\documentclass[a4paper,11pt]{article}
\pdfoutput=1 % if your are submitting a pdflatex (i.e. if you have
\usepackage{jheppub}
\usepackage{hyperref}
\usepackage{amsfonts}
\usepackage{amsmath}
\usepackage{amssymb}
\usepackage[english]{babel}
\usepackage{graphicx}
\usepackage{epsfig}
\usepackage{bm}
\usepackage{longtable}
\usepackage{verbatim}
\usepackage{longtable}
\usepackage[utf8]{inputenc}
\usepackage{color}

\def\vev#1{\left\langle #1\right\rangle}

\definecolor{greenLinks}{rgb}{0, 0.6, 0} 
\definecolor{blueLinks}{rgb}{0, 0, 0.6}
\definecolor{redLinks}{rgb}{0.6, 0, 0}
\definecolor{tempText}{rgb}{0.55, 0.10,0.67}
\definecolor{eprintLinks}{rgb}{0.4, 0.4, 0.4}
\definecolor{journalLinks}{rgb}{0.6, 0, 0}
\newcommand {\ignore}[1]{}

\newcommand{\sm}{{standard model }}

\def\lsim{\mathrel{\rlap{\lower4pt\hbox{\hskip1pt$\sim$}}
    \raise1pt\hbox{$<$}}}
\def\gsim{\mathrel{\rlap{\lower4pt\hbox{\hskip1pt$\sim$}}
    \raise1pt\hbox{$>$}}}

\def\3211{$\mathrm{SU(3) \otimes SU(2)_L \otimes U(1)_R \otimes U(1)_{B-L}}$ }
\def\321{$\mathrm{SU(3) \otimes SU(2) \otimes U(1)}$ }
\def\422{$\mathrm{SU(4) \otimes SU(2) \otimes SU(2)_R}$ }

\def \nbb {$\rm 0\nu \beta \beta $ }

\newcommand{\mathsym}[1]{{}}

\title{Predictive Pati-Salam theory of fermion masses and mixing}

\author[a]{A. E. C\'arcamo Hern\'andez}
\author[a]{Sergey Kovalenko}
\author[b]{Jos\'e W.F. Valle}
\author[b]{C.A. Vaquera-Araujo}

\affiliation[a]{Universidad T\'{e}cnica Federico Santa Mar\'{\i}a\\
and Centro Cient\'{\i}fico-Tecnol\'{o}gico de Valpara\'{\i}so\\
Casilla 110-V, Valpara\'{\i}so, Chile}
\affiliation[b]{AHEP Group, Institut de F\'{i}sica Corpuscular --
  C.S.I.C./Universitat de Val\`{e}ncia, Parc Cient\'ific de Paterna.\\
 C/ Catedr\'atico Jos\'e Beltr\'an, 2 E-46980 Paterna (Valencia) - SPAIN}

\emailAdd{antonio.carcamo@usm.cl}
\emailAdd{sergey.kovalenko@usm.cl}
\emailAdd{valle@ific.uv.es}
\emailAdd{vaquera@ific.uv.es}

\abstract{We propose a Pati-Salam extension of the {standard model} incorporating a
flavor symmetry based on the $\Delta \left( 27\right)$ group. The theory
realizes a realistic Froggatt-Nielsen picture of quark mixing and a
predictive pattern of neutrino oscillations. We find that, for normal
neutrino mass ordering, the atmospheric angle must lie in the higher octant,
CP must be violated in oscillations, and there is a lower bound for the \nbb %
decay rate. For the case of inverted mass ordering, we find that the lower
atmospheric octant is preferred, and that CP can be conserved in
oscillations. Neutrino masses arise from a low-scale seesaw mechanism, whose
messengers can be produced by a $Z^{\prime }$ portal at the LHC.}

\begin{document}
\maketitle
\flushbottom

\section{Introduction}

\label{sec:introduction}

Apart from the discovery of neutrino oscillations~\cite%
{Kajita:2016cak,McDonald:2016ixn}, no other laboratory evidence for physics
beyond the standard model has been so far unambiguously confirmed. Both the
origin of neutrino mass itself, as well as the understanding of the mixing
pattern, require an explanation from \textit{first principles}. Moreover,
there is a variety of other motivations for having beyond the {standard model%
} physics~\cite{Valle:2015pba}. One of these is the pursuit of a dynamical
explanation for the origin of parity violation in the weak interaction,
whose basic V-A nature is put in by hand in the formulation of the standard
model. With this in mind here we propose a flavored~\cite%
{Ishimori:2010au,Morisi:2012fg,King:2013eh} Pati-Salam~\cite{pati:1974yy}
extension of the {standard model}, addressing both the dynamical origin of
the V-A nature of the weak force, as well as the related origin of neutrino
mass. In addition, as we will see, the model can shed light upon the flavor
problem and make predictions. The main features of our model include:

\begin{itemize}
\item adequate implementation of $\Delta \left( 27\right) $ flavor symmetry
in the Pati-Salam framework and symmetry breaking

\item consistent low-scale left-right symmetric seesaw mechanism for
neutrinos~\cite%
{Mohapatra:1986bd,Akhmedov:1995ip,Akhmedov:1995vm,Malinsky:2005bi,Perez:2013osa}

\item predictive pattern of neutrino mixing summarized in Figs.~\ref{NHplot}
and \ref{IHplot}

\item realistic pattern of quark mixing, yielding a Froggatt-Nielsen~\cite%
{Froggatt:1978nt}-like picture of the CKM matrix

\item lower bound for the \nbb decay rate in Fig.~\ref{mee}

\end{itemize}

We note also that the model has a low-scale $Z^{\prime }$ portal through
which the TeV scale messengers $S_{a}$ can be pair-produced in Drell-Yan
collisions at the LHC~\cite%
{Deppisch:2013cya,AguilarSaavedra:2012fu,Das:2012ii}. In addition, our model
realizes a universal seesaw mechanism \cite{Davidson:1987mh} for the down
type quarks as well as the charged leptons, mediated by TeV scale exotic
fermions. The latter should potentially lead to other phenomenological
effects in the quark sector as well as lepton flavour violation effects.

\section{The model}

\label{sec:model}

The model under consideration is based on the Pati-Salam gauge symmetry 
$SU(4)_{C}\otimes SU\left( 2\right)_{L}\otimes SU\left( 2\right)_{R}$,
supplemented by the $\Delta \left( 27\right) \otimes Z_{4}\otimes Z_{16}$
discrete family symmetry group. The fermion transformation properties under
the Pati-Salam group are 
\begin{align}
\overline{F}_{iL} &\sim \left( \overline{\mathbf{4}},\mathbf{2,1}\right) , & 
F_{iR}&\sim \left( \mathbf{4},\mathbf{1,2}\right) , & S_{i}&\sim \left( 
\mathbf{1},\mathbf{1,1}\right) ,  \notag \\
\overline{\Psi }_{iL} &\sim \left( \overline{\mathbf{4}},\mathbf{1,1}\right)
, & \Psi_{iR}&\sim \left( \mathbf{4},\mathbf{1,1}\right) , & i&=1,2,3,
\end{align}%
where the subscript refers to fermion families. More explicitly, the {%
standard model } fermions are written in component form as 
\begin{equation}
\overline{F}_{iL}=\left( 
\begin{array}{cccc}
\overline{u}_{iL} & \overline{u}_{iL} & \overline{u}_{iL} & \overline{\nu }%
_{iL} \\ 
\overline{d}_{iL} & \overline{d}_{iL} & \overline{d}_{iL} & \overline{l}_{iL}%
\end{array}%
\right) ^{T},\hspace{1.5cm}F_{iR}=\left( 
\begin{array}{cccc}
u_{iR} & u_{iR} & u_{iR} & \nu_{iR} \\ 
d_{iR} & d_{iR} & d_{iR} & l_{iR}%
\end{array}%
\right).
\end{equation}
Notice that we have extended the fermion sector of the original
Pati-Salam model~\cite{pati:1974yy} by introducing three fermion
singlets $S_{i}$, in order to implement inverse and/or linear seesaw
mechanisms for the generation of light active neutrino
masses~\cite{Mohapatra:1986bd,Akhmedov:1995ip,Akhmedov:1995vm,Malinsky:2005bi,Perez:2013osa}.
In addition, we have introduced vector-like fermions $\overline{\Psi
}%
_{iL} $ and $\Psi_{iR}$ so as to generate the {standard model }
down-type quark and charged lepton masses, from a universal seesaw
mechanism.

The particle content and gauge symmetry assignments are summarized in table~%
\ref{tab:gqn}. 
\begin{table}[ht]
\centering
\begin{tabular}{|c||c|c|c|c|c||c|c|c|c|c|c||c|c|c|c|c|}
\hline\hline
Field & $\overline{F}_{iL}$ & $F_{iR}$ & $\overline{\Psi}_{iL}$ & $\Psi
_{iR}$ & $S_{i}$ & $\chi_{L}$ & $\chi_{R}$ & $\Phi_{j}$ & $\Sigma$ & $\phi
_{k}$ & $\varphi_{l}$ & $\sigma$ & $\rho_{k}$ & $\eta_{k}$ & $\tau_{k}$ & $%
\xi_{k}$ \\ \hline
$SU(4)_{C}$ & $\overline{\mathbf{4}}$ & $\mathbf{4}$ & $\overline{\mathbf{4}}
$ & $\mathbf{4}$ & $\mathbf{1}$ & $\mathbf{4}$ & $\mathbf{4}$ & $\mathbf{1}$
& $\mathbf{15}$ & $\mathbf{1}$ & $\mathbf{1}$ & $\mathbf{1}$ & $\mathbf{1}$
& $\mathbf{1}$ & $\mathbf{1}$ & $\mathbf{1}$ \\ \hline
$SU\left( 2\right)_{L}$ & $\mathbf{2}$ & $\mathbf{1}$ & $\mathbf{1}$ & $%
\mathbf{1}$ & $\mathbf{1}$ & $\mathbf{2}$ & $\mathbf{1}$ & $\mathbf{2}$ & $%
\mathbf{1}$ & $\mathbf{2}$ & $\mathbf{1}$ & $\mathbf{1}$ & $\mathbf{1}$ & $%
\mathbf{1}$ & $\mathbf{1}$ & $\mathbf{1}$ \\ \hline
$SU\left( 2\right)_{R}$ & $\mathbf{1}$ & $\mathbf{2}$ & $\mathbf{1}$ & $%
\mathbf{1}$ & $\mathbf{1}$ & $\mathbf{1}$ & $\mathbf{2}$ & $\mathbf{2}$ & $%
\mathbf{1}$ & $\mathbf{1}$ & $\mathbf{2}$ & $\mathbf{1}$ & $\mathbf{1}$ & $%
\mathbf{1}$ & $\mathbf{1}$ & $\mathbf{1}$ \\ \hline\hline
\end{tabular}
\caption{Particle content and transformation properties under the gauge
symmetry.}
\label{tab:gqn}
\end{table}

The fermion assignments under the flavor symmetry group $\Delta \left(
27\right) \otimes Z_{4}\otimes Z_{16}$ are: 
\begin{align}
\overline{F}_{1L} &\sim \left( \mathbf{1}_{\mathbf{0,0}},1,i\right) , & 
\overline{F}_{2L}&\sim \left( \mathbf{1}_{\mathbf{2,0}},i,i^{\frac{1}{2}%
}\right) , & \overline{F}_{3L}&\sim \left( \mathbf{1}_{\mathbf{1,0}%
},1,1\right) ,  \notag \\
F_{1R} &\sim \left( \mathbf{1}_{\mathbf{0,0}},1,i\right) , & F_{2R}&\sim
\left( \mathbf{1}_{\mathbf{1,0}},i,i^{\frac{1}{2}}\right) , & F_{3R}&\sim
\left( \mathbf{1}_{\mathbf{2,0}},1,1\right) ,  \notag \\
\overline{\Psi }_{1L} &\sim \left( \mathbf{1}_{\mathbf{0,0}},1,1\right) & 
\overline{\Psi }_{2L}&\sim \left( \mathbf{1}_{\mathbf{2,0}},1,1\right) , & 
\overline{\Psi }_{3L} &\sim \left( \mathbf{1}_{\mathbf{1,0}},1,1\right) 
\notag \\
\Psi_{1R} &\sim \left( \mathbf{1}_{\mathbf{0,0}},1,1\right) , & \Psi
_{2R}&\sim \left( \mathbf{1}_{\mathbf{1,0}},1,1\right) , & \Psi_{3R}&\sim
\left( \mathbf{1}_{\mathbf{2,0}},1,1\right),  \notag \\
& & S & \sim \left( \mathbf{3,}i,1\right), & 
\end{align}
where the numbers in boldface stand for the dimensions of the $\Delta \left(
27\right) $ irreducible representations, and we have defined $S \equiv\left(
S_{1},S_{2},S_{3}\right)$. 
Notice that the three Pati-Salam singlet fermions are
grouped into a $\Delta \left( 27\right) $ triplet $S$, whereas the remaining
fermions are assigned to singlet representations of $\Delta \left( 27\right) 
$. These fermion flavor symmetry transformation properties are summarized in
table~\ref{tab:ffqn}.  A brief summary of the properties of $\Delta \left( 27\right) $ group and its irreducible representations is given in Appendix \ref{sec:product-rules-delta}.
\begin{table}[!h]
\centering
\begin{tabular}{|c||c|c|c|c|c|c|c|c|c|c|c|c|c|}
\hline\hline
Field & $\overline{F}_{1L}$ & $\overline{F}_{2L}$ & $\overline{F}_{3L}$ & $%
F_{1R}$ & $F_{2R}$ & $F_{3R}$ & $\overline{\Psi }_{1L}$ & $\overline{\Psi }%
_{2L}$ & $\overline{\Psi }_{3L}$ & $\Psi_{1R}$ & $\Psi_{2R}$ & $\Psi_{3R}$
& $S$ \\ \hline
$\Delta(27)$ & $\mathbf{1}_{\mathbf{0,0}}$ & $\mathbf{1}_{\mathbf{2,0}}$ & $%
\mathbf{1}_{\mathbf{1,0}}$ & $\mathbf{1}_{\mathbf{0,0}}$ & $\mathbf{1}_{%
\mathbf{1,0}}$ & $\mathbf{1}_{\mathbf{2,0}}$ & $\mathbf{1}_{\mathbf{0,0}}$ & 
$\mathbf{1}_{\mathbf{2,0}}$ & $\mathbf{1}_{\mathbf{1,0}}$ & $\mathbf{1}_{%
\mathbf{0,0}}$ & $\mathbf{1}_{\mathbf{1,0}}$ & $\mathbf{1}_{\mathbf{2,0}}$ & 
$\mathbf{3}$ \\ \hline
$Z_{4}$ & $1$ & $i$ & $1$ & $1$ & $i$ & $1$ & $1$ & $1$ & $1$ & $1$ & $1$ & $%
1$ & $i$ \\ \hline
$Z_{16}$ & $i$ & $i^{\frac{1}{2}}$ & $1$ & $i$ & $i^{\frac{1}{2}}$ & $1$ & $%
1 $ & $1$ & $1$ & $1$ & $1$ & $1$ & $1$ \\ \hline\hline
\end{tabular}%
\caption{Transformation properties of the leptons and quarks under the
flavor symmetry $\Delta(27) \otimes Z_4\otimes Z_4^\prime$.}
\label{tab:ffqn}
\end{table}

%\color{cyan}

We now turn to the main features of the scalar sector. The Higgs fields of the model are assumed to transform under the Pati-Salam gauge group as 
\begin{align}
\chi_{R} &\sim \left( \mathbf{4},\mathbf{1,2}\right) , & \chi_{L}&\sim
\left( \mathbf{4},\mathbf{2,1}\right) , & \Phi_{j}&\sim \left( \mathbf{1},%
\mathbf{2,2}\right) , & \Sigma &\sim \left( \mathbf{15},\mathbf{1,1}\right) ,
& j&=1,2,  \notag \\
\phi_{k} &\sim \left( \mathbf{1},\mathbf{2,1}\right) , & \varphi_{l}&\sim
\left( \mathbf{1},\mathbf{1,2}\right) , & \sigma &\sim \left( \mathbf{1},%
\mathbf{1,1}\right) , & \rho_{k}&\sim \left( \mathbf{1},\mathbf{1,1}\right)
, & l&=1,2,3,4,5,  \notag \\
\eta_{k} &\sim \left( \mathbf{1},\mathbf{1,1}\right) , & \tau_{k}&\sim
\left( \mathbf{1},\mathbf{1,1}\right) , & \xi_{k}&\sim \left( \mathbf{1},%
\mathbf{1,1}\right) , & & & k&=1,2,3,
\end{align}
and the scalars $\chi_{L,R}$, $\Phi_{k} $ and $\Sigma $ develop vacuum
expectation values (VEVs) of the form 
\begin{equation}
\left\langle \chi_{L,R}\right\rangle =\left( 
\begin{array}{cccc}
0 & 0 & 0 & v_{L,R} \\ 
0 & 0 & 0 & 0%
\end{array}%
\right) ,\hspace{1.5cm}\left\langle \Phi_{j}\right\rangle =\left( 
\begin{array}{cc}
v_{1}^{\left( j\right) } & 0 \\ 
0 & v_{2}^{\left( j\right) }%
\end{array}%
\right) ,\hspace{1.5cm}\left\langle \Sigma \right\rangle =v_{\Sigma }T^{15},
\end{equation}%
with $T^{15}=\frac{1}{2\sqrt{6}}\mathrm{diag}\left( 1,1,1,-3\right) $.

%\color{red}

The transformation properties of the scalar fields under the $\Delta \left(
27\right) \otimes Z_{4}\otimes Z_{16}$ discrete group are given as follows: 
\begin{align}
\chi_{R} &\sim \left( \mathbf{1}_{\mathbf{0,0}}\mathbf{,}i,1\right) , & 
\chi_{L}&\sim \left( \mathbf{1}_{\mathbf{0,0}}\mathbf{,-}i,1\right) , & 
\Phi_{1}&\sim \left( \mathbf{1}_{\mathbf{0,0}},1,1\right) , & \Phi
_{2}&\sim \left( \mathbf{1}_{\mathbf{0,0}},-1,1\right) ,  \notag \\
\Sigma &\sim \left( \mathbf{1}_{\mathbf{0,0}},1,e^{\frac{3\pi i}{4}}\right) ,
& \phi_{1}&\sim \left( \mathbf{1}_{\mathbf{0,0}},1,e^{-\frac{2\pi i}{8}%
}\right) , & \phi_{2}&\sim \left( \mathbf{1}_{\mathbf{0,0}},-i,e^{-\frac{%
\pi i}{8}}\right) , & \phi_{3}&\sim \left( \mathbf{1}_{\mathbf{0,0}%
},1,1\right) ,  \notag \\
\varphi_{1} &\sim \left( \mathbf{1}_{\mathbf{0,0}},1,e^{-\frac{2\pi i}{8}%
}\right) , & \varphi_{2}&\sim \left( \mathbf{1}_{\mathbf{0,0}},-i,e^{-\frac{%
\pi i}{8}}\right) , & \varphi_{3}&\sim \left( \mathbf{1}_{\mathbf{0,0}%
},1,1\right) , & \varphi_{4}&\sim \left( \mathbf{1}_{\mathbf{2,0}},-i,e^{-%
\frac{\pi i}{8}}\right) ,  \notag \\
\varphi_{5} &\sim \left( \mathbf{1}_{\mathbf{2,0}},1,e^{\frac{\pi i}{8}%
}\right) , & \sigma &\sim \left( \mathbf{1}_{\mathbf{0,0}},1,e^{-\frac{\pi i%
}{8}}\right) , & \rho &\sim \left( \overline{\mathbf{3}},1,i\right) & \eta &
\sim \left( \overline{\mathbf{3}},-i,i^{\frac{1}{2}}\right),  \notag \\
& & \tau &\sim \left( \overline{\mathbf{3}},1,1\right) , & \xi &\sim \left( 
\overline{\mathbf{3}},-1,1\right) & 
\end{align}
where we have set 
\begin{align}
\rho &\equiv\left( \rho_{1},\rho_{2},\rho_{3}\right) & \eta &=\left( \eta
_{1},\eta_{2},\eta_{3}\right) & \tau &\equiv\left( \tau_{1},\tau
_{2},\tau_{3}\right) & \xi &\equiv\left( \xi_{1},\xi_{2},\xi_{3}\right) .
\end{align}

These flavor symmetry scalar transformation properties are summarized in
table~\ref{tab:sfqn}. 
\begin{table}[ht]
\centering
\scalebox{0.8}{%
\begin{tabular}{|c||c|c|c|c|c|c|c|c|c|c|c|c|c||c|c|c|c|c|}
\hline\hline
Field & $\chi_{L}$ & $\chi_{R}$ & $\Phi_{1}$ & $\Phi_{2}$ & $\Sigma$ & $%
\phi_{1}$ & $\phi_{2}$ & $\phi_{3}$ & $\varphi_{1}$ & $\varphi_{2}$ & $%
\varphi_{3}$ & $\varphi_{4}$ & $\varphi_{5}$ & $\sigma$ & $\rho$ & $\eta$
& $\tau$ & $\xi$ \\ \hline
$\Delta(27)$ & $\mathbf{1}_{\mathbf{0,0}}$ & $\mathbf{1}_{\mathbf{0,0}}$ & $%
\mathbf{1}_{\mathbf{0,0}}$ & $\mathbf{1}_{\mathbf{0,0}}$ & $\mathbf{1}_{%
\mathbf{0,0}}$ & $\mathbf{1}_{\mathbf{0,0}}$ & $\mathbf{1}_{\mathbf{0,0}}$ & 
$\mathbf{1}_{\mathbf{0,0}}$ & $\mathbf{1}_{\mathbf{0,0}}$ & $\mathbf{1}_{%
\mathbf{0,0}}$ & $\mathbf{1}_{\mathbf{0,0}}$ & $\mathbf{1}_{\mathbf{2,0}}$ & 
$\mathbf{1}_{\mathbf{2,0}}$ & $\mathbf{1}_{\mathbf{0,0}}$ & $\overline{%
\mathbf{3}}$ & $\overline{\mathbf{3}}$ & $\overline{\mathbf{3}}$ & $%
\overline{\mathbf{3}}$ \\ \hline
$Z_{4}$ & $-i$ & $i$ & $1$ & $-1$ & $1$ & $1$ & $-i$ & $1$ & $1$ & $-i$ & $1$
& $-i$ & $1$ & $1$ & $1$ & $-i$ & $1$ & $-1$ \\ \hline
$Z_{16}$ & $1$ & $1$ & $1$ & $1$ & $i^{\frac{3}{2}}$ & $i^{-\frac{1}{2}}$ & $i^{-\frac{1}{4}}$ & $1$ & $i^{-\frac{1}{2}}$ & $i^{-\frac{1}{4}}$ & $1$ & $i^{-\frac{1}{4}}$ & $i^{\frac{1}{4}}$ & $i^{-\frac{1}{4}}$ & $i$ & $i^{\frac{1}{2}}$ & $1$ & $1$ \\ \hline\hline
\end{tabular}
}%
\caption{Transformation properties of the scalars under the flavor symmetry $%
\Delta(27) \otimes Z_4\otimes Z_4^\prime$.}
\label{tab:sfqn}
\end{table}

With the above particle content and transformation properties the following
relevant Yukawa terms arise: 
\begin{eqnarray}
-\mathcal{L}_{Y} &=&y_{1}\overline{F}_{1L}\Phi_{1}F_{1R}\frac{\sigma ^{8}}{%
\Lambda ^{8}}+y_{2}\overline{F}_{2L}\Phi_{2}F_{2R}\frac{\sigma ^{4}}{%
\Lambda ^{4}}+y_{3}\overline{F}_{3L}\Phi_{1}F_{3R}  
\label{eq:Yukawa-1} \\
&&+\frac{\alpha_{1}}{\Lambda }\overline{F}_{1L}\chi_{L}\left( S\rho
\right)_{\mathbf{1}_{\mathbf{0,0}}}\frac{\sigma ^{8}}{\Lambda ^{8}}+\frac{%
\alpha_{2}}{\Lambda }\overline{F}_{2L}\chi_{L}\left( S\eta \right)_{%
\mathbf{1}_{\mathbf{1,0}}}\frac{\sigma ^{4}}{\Lambda ^{4}}+\frac{\alpha_{3}%
}{\Lambda }\overline{F}_{3L}\chi_{L}\left( S\tau \right)_{\mathbf{1}_{%
\mathbf{2,0}}}  \notag \\
&&+\frac{\beta_{1}}{\Lambda }F_{1R}\chi_{R}^{\dagger }\left( S\rho \right)
_{\mathbf{1}_{\mathbf{0,0}}}\frac{\sigma ^{8}}{\Lambda ^{8}}+\frac{\beta_{2}%
}{\Lambda }F_{2R}\chi_{R}^{\dagger }\left( S\eta \right)_{\mathbf{1}_{%
\mathbf{2,0}}}\frac{\sigma ^{4}}{\Lambda ^{4}}+\frac{\beta_{3}}{\Lambda }%
F_{3R}\chi_{R}^{\dagger }\left( S\tau \right)_{\mathbf{1}_{\mathbf{1,0}}} 
\notag \\
&&+\kappa_{1}\overline{F}_{1L}\phi_{1}\Psi_{1R}\frac{\sigma ^{2}}{\Lambda
^{2}}+\kappa_{2}\overline{F}_{2L}\phi_{2}\Psi_{2R}\frac{\sigma }{\Lambda }%
+\kappa_{3}\overline{F}_{3L}\phi_{3}\Psi_{3R}  \notag \\
&&+\gamma_{1}\overline{\Psi }_{1L}\varphi_{1}F_{1R}\frac{\sigma ^{2}}{%
\Lambda ^{2}}+\gamma_{2}\overline{\Psi }_{2L}\varphi_{2}F_{2R}\frac{\sigma 
}{\Lambda }+\gamma_{3}\overline{\Psi }_{3L}\varphi_{3}F_{3R}  \notag \\
&&+\gamma_{4}\overline{\Psi }_{1L}\varphi_{4}F_{2R}\frac{\sigma }{\Lambda }%
+\gamma_{5}\overline{\Psi }_{1L}\varphi_{5}^{\ast }F_{3R}\frac{\sigma
^{\ast }}{\Lambda }+\gamma_{6}\overline{\Psi }_{2L}\varphi_{5}F_{3R}\frac{%
\sigma }{\Lambda }  \notag \\
&&+\sum\limits_{i=1}^{3}\left[ A_{i}\overline{\Psi }_{iL}\Psi_{iR}+a_{i}%
\overline{\Psi }_{riL}\left( \Sigma \right)_{s}^{r}\Psi_{iR}^{s}\frac{%
\sigma ^{6}}{\Lambda ^{6}}\right] +
\lambda_{1}\left( \overline{S} S^{C}\right)_{\mathbf{3}_{S_{1}}}\xi +\lambda_{2}\left( \overline{S}%
S^{C}\right)_{\mathbf{3}_{S_{2}}}\xi +\mathrm{h.c.}, \notag
\end{eqnarray}%
where $r$, $s$ are $SU\left( 4\right) $ indices and $y_{i}$, $\alpha_{i}$, $%
\beta_{i}$, $\kappa_{i}$, $a_{i}$ ($i=1,2,3$), $\gamma_{m}$ ($%
m=1,2,\cdots ,6$) and $\lambda_{j}$\ ($j=1,2$) are $\mathcal{O}(1)$
dimensionless couplings.  For an explanation of the $\Delta(27)$ notation used in 
the $\alpha,\beta$ and $\lambda$-terms, see Appendix 
\ref{sec:product-rules-delta}.  It is noteworthy that the lightest of the physical
neutral scalar states of $\left( \Phi_{j}\right)_{11}$, $\left( \Phi
_{j}\right)_{22}$, $\phi_{i}$ should be interpreted as the SM-like 125 GeV
Higgs recently found at the LHC. Furthermore, our model at low energies
corresponds to a seven Higgs doublet model with five scalar singlets 
(these scalar singlets come from $\varphi_{l}$). 
As we will show in section \ref%
{quarkmassesandmixing}, the top quark mass only arises from $\left( \Phi
_{1}\right)_{11}$. Consequently, the dominant contribution to the SM-like
125 GeV Higgs mainly arises from $\left( \Phi_{1}\right)_{11}$. We note
also that the scalar potential of our model has many free parameters, so
that we can assume the remaining scalars to be heavy and outside the LHC
reach. Moreover, one can suppress the loop effects of the heavy scalars
contributing to precision observables, by making an appropriate choice of
the free parameters in the scalar potential. These adjustments do not affect
the physical observables in the quark and lepton sectors, which are
determined mainly by the Yukawa couplings.

%\color{black}

The full symmetry group $\mathcal{G}$ exhibits the following spontaneous
breaking pattern: 
\begin{eqnarray}
&\mathcal{G}=SU(4)_{C}\otimes SU\left( 2\right)_{L}\otimes SU\left(
2\right)_{R}\otimes \Delta \left( 27\right) \otimes Z_{4}\otimes Z_{16} & 
 \\
& \langle \Sigma\rangle\sim \Lambda_{PS} &  \notag \\
&\Downarrow&\notag\\
&SU(3)_{C}\otimes SU\left( 2\right)_{L}\otimes SU\left( 2\right)
_{R}\otimes U\left( 1\right)_{B-L}&  \\
&\langle \chi_{R}\rangle \sim\  v_{R}  \ \   &\notag \\
&\Downarrow& \notag\\
&SU(3)_{C}\otimes SU\left( 2\right)_{L}\otimes U\left( 1\right)_{Y} & 
\\
&\langle \Phi_{{1,2}}\rangle \sim\ v \ \ \ \ &  \notag \\
&\Downarrow&\notag \\
&SU(3)_{C}\otimes U\left( 1\right)_{Q}.&
\end{eqnarray}
Here $v=246$ GeV is the electroweak symmetry breaking scale, and we
assume that the Pati-Salam gauge symmetry is broken at the scale
$\Lambda_{PS} \mathrel{\rlap{\lower4pt\hbox{\hskip1pt$\sim$}}
  \raise1pt\hbox{$\gsim$}}10^{6}$ GeV. This restriction follows from
the experimental limit on the branching ratio for the rare neutral
meson decays, such as $B^0 \to l_i^{\pm} l_j^{\mp}$, mediated by the
vector leptoquarks, as discussed in
Refs.~\cite{Valencia:1994cj,Smirnov:2007hv}. Furthermore, it is worth
mentioning that Pati-Salam models with a quark-lepton unification
scale of about $%
\mathrel{\rlap{\lower4pt\hbox{\hskip1pt$\sim$}} \raise1pt\hbox{$>$}}
10^{6}$ GeV can fulfill gauge coupling unification
\cite{Hartmann:2014fya}. A comprehensive study of gauge coupling
unification in models that include all possible chains of Pati-Salam
symmetry breaking in both supersymmetric and non-supersymmetric
scenarios has been given in Ref. \cite{Hartmann:2014fya}.

\section{Understanding the model setup}
\label{sec:underst-model-setup}

In this section we try to motivate in more detail our choice for the model
content and the transformation properties. First note that the Pati-Salam gauge
symmetry $SU(4)_{C}\otimes SU\left( 2\right)_{L}\otimes SU\left( 2\right)
_{R}$ breaks down to the conventional left-right symmetry $SU(3)_{C}\otimes
SU\left( 2\right)_{L}\otimes SU\left( 2\right)_{R}\otimes U\left( 1\right)
_{B-L}$ by the VEV of the scalar field $\Sigma $, at the scale $\Lambda_{PS}%
\mathrel{\rlap{\lower4pt\hbox{\hskip1pt$\sim$}}
  \raise1pt\hbox{$>$}}10^{6}$ GeV. The next symmetry breaking step is
triggered by $\chi_{R}$, whose VEV is assumed to be in the few TeV scale,
playing an important role in implementing the low-scale seesaw neutrino mass
generation~\cite{Akhmedov:1995ip,Akhmedov:1995vm}. The breaking of the
electroweak gauge group $SU(3)_{C}\otimes SU\left( 2\right)_{L}\otimes
U\left( 1\right)_{Y}$ is triggered by the scalar fields $\Phi_{j}$, which
acquire vacuum expectation values at the electroweak symmetry breaking scale 
$v=246$ GeV.

Besides, note that the presence of the scalar field $\Sigma $
transforming as the adjoint representation of $SU\left( 4\right) $ is
also crucial in the implementation of the Universal Seesaw mechanism
for down-type quarks and charged leptons, mediated by exotic
fermions. This scalar field $\Sigma $ acquires a VEV at the scale
$\Lambda_{PS}$, so that an insertion
$\frac{\sigma ^{6}}{\Lambda ^{6}}$ in its corresponding Yukawa term
generates a TeV scale contribution to the exotic charged lepton and
exotic down-type quark masses. The charge of $\Sigma $ under the
$Z_{16}$ discrete group ensures that its different contributions to the charged leptons and down type quark masses will be comparable to the ones arising from the $A_{i}\overline{\Psi }_{iL}\Psi_{iR}$ ($i=1,2,3$) mass terms (which contribute equally to the down type quark and charged lepton masses). It is worth mentioning that we are assuming $A_{i}\approx\mathcal{O}(1)$ TeV. Let us note that the inclusion of the scalar field $\Sigma $ is necessary to guarantee that the resulting
down-type quark masses are different from the charged lepton masses, as it will be shown in Section \ref{quarkmassesandmixing}.

Notice that the
scalars $\phi_{i}$ and $\varphi_{l}$ are needed to generate the
mixing between the {standard model } charged leptons and down-type
quarks with their exotic siblings, so as to implement the Universal
Seesaw mechanism that gives rise to realistic masses for the standard
model charged fermions.
The scalar fields 
$\chi_{R}, \chi_{L}$ and $\xi_{i}$ 
have Yukawa
terms necessary for the implementation of the inverse and linear
seesaw mechanisms, so as to generate the light active neutrino masses.
This requires also that VEVs of $\xi_{i}$ are much smaller than the
electroweak symmetry breaking scale.

The scalar fields $\rho_{i}$ , $\eta_{i}$ and $\tau_{i}$ are needed
to generate the diagonal $3\times 3$ blocks that include the mixing of
the neutrino states contained in $\overline{F}_{iL} $ and $F_{iR}$
with the singlet neutrinos $S_{i}$ ($i=1,2,3$), thus avoiding the
transmission of the strong hierarchy in the up mass matrix to the
light active neutrino mass matrix. Furthermore, the scalar field
$\sigma $, charged under the $Z_{16}$ discrete group is need to
generate the observed SM charged fermion mass and quark mixing
hierarchy. In order to relate the quark masses with the quark mixing
parameters, we assume that the scalar field $\sigma $ acquires a VEV
equal to $\lambda \Lambda $, where $\lambda =0.225$ is one of the
Wolfenstein parameters and $\Lambda $ is the cutoff of our model. In
summary, the set of VEVs of the scalar fields is assumed to satisfy
the following hierarchy:
\begin{equation}
v_{L}\ll v_{\xi }\ll v_{1}^{\left( j\right) }\sim v_{\phi_{i}}\sim
v_{\varphi_{l}}\sim v\ll v_{R}\ll v_{\rho }=v_{\eta }=v_{\tau }\sim \Lambda
_{PS}\sim v_{\sigma }=\lambda \Lambda ,  \label{VEVhierarchy}
\end{equation}
where  $\langle\chi_{L,R}\rangle= v_{L,R}$.
We now comment on the possible VEV patterns for the $\Delta (27)$
scalar triplets $\rho $, $\eta $, $\tau $ and $\xi $. Since the VEVs
of the $\Delta (27)$ scalar triplets satisfy the hierarchy
$v_{\xi }\ll v_{\rho }=v_{\eta }=v_{\tau }\sim \Lambda_{PS}$, the
mixing angles of $\xi $ with $\rho $, $\eta $ and $\tau $ are very
tiny since they are suppressed by the ratios of their VEVs, and
consequently the method of recursive expansion proposed in
Ref. \cite{Grimus:2000vj} can be used for the analysis of the model scalar potential. 
In this scenario, the scalar potential for the $\Delta (27)$ scalar
triplet $\xi $ can be treated independently from the scalar potential
for the $\Delta (27)$ scalar triplets $\rho $, $\eta $, $\tau $, as
shown in detail in the appendices \ref{sec:scalarpotential1triplet}
and \ref{sec:scalarpotential3triplets}. One can see that the following
VEV patterns for the $\Delta (27)$ scalar triplets are consistent with
the scalar potential minimization equations for a large region of
parameter space
\begin{eqnarray}
  \vev{ \rho} &=&v_{\rho }\left( 1,0,0\right) ,\hspace{1.5cm}
  \vev {\eta} =v_{\eta }\left( 0,1,0\right) ,\hspace{1.5cm}
  \vev {\tau} =v_{\tau }\left( 0,0,1\right) ,  \notag
  \\
  \vev{ \xi} &=&\frac{v_{\xi }}{\sqrt{2+r^2}}\left(r,e^{-i\psi },e^{i\psi }\right) .  
\label{VEVpattern}
\end{eqnarray}

We now turn our attention on the role of each discrete group factor of
our model.
As will be seen in Sects. \ref{quarkmassesandmixing} and
\ref{leptonmassesandmixing} the $\Delta \left( 27\right) $ discrete
group is crucial for the predictivity of our model, giving rise to
viable textures for the fermion masses and mixings.
Notice that the $\Delta (27)$ discrete group is a non
trivial group of the type $\Delta (3n^{2})$ for $n=3$, isomorphic to the
semi-direct product group
$(Z_{3}^{\prime }\times Z_{3}^{\prime \prime })\rtimes Z_{3}$
\cite{Ishimori:2010au}. Recently, this group has been used in
multi-Higgs doublet models \cite{Bhattacharyya:2012pi}, $SO(10)$
models \cite{Bjorkeroth:2015uou}, warped extra dimensional models
\cite{Chen:2015jta} and models based on the
$SU(3)_{C}\otimes SU(3)_{L}\otimes U(1)_{X}$ gauge symmetry
\cite{Vien:2016tmh,Hernandez:2016eod}. 
We introduce the $Z_{16}$ discrete group, since it is the smallest
cyclic symmetry that allows the Yukawa term
$\overline{ F}_{1L}\Phi_{1}F_{1R}\frac{\sigma ^{8}}{\Lambda ^{8}}$ of
dimension twelve from a $\frac{\sigma ^{8}}{\Lambda ^{8}}$ insertion
on the $\overline{F}_{1L}\Phi_{1}F_{1R}$ operator.
This leads to the required $\lambda ^{8}$ suppression needed to
naturally explain the smallness of the up quark mass.  The $Z_{16}$
group has been recently shown to be useful for explaining the observed
SM charged fermion mass and quark mixing hierarchy, in the framework
of a $SU(3)_{C}\otimes SU(3)_{L}\otimes U(1)_{X}$ models based on the
$A_{4}$ and $S_{3}$ family symmetries
\cite{Hernandez:2015tna,CarcamoHernandez:2017kra,Hernandez:2013hea}. 
As we will see in the next section, in our model the CKM matrix arises
from the down-type quark sector. In order to get the correct hierarchy
in the entries of the quark mass matrices yielding a realistic pattern
of quark masses and mixing angles, we use a $Z_{4}$ discrete symmetry
and the scalar bidoublets $\Phi_{j}$ ($j=1,2$), one neutral and the
another charged under $Z_{4}$. 
This group was previously used in some other flavor models and proved to be helpful, in particular, in the context of Grand Unification \cite{Emmanuel-Costa:2013gia,Arbelaez:2015toa}, models with extended $SU(3)_{C}\otimes SU(3)_{L}\otimes U(1)_{X}$ gauge symmetry \cite{Hernandez:2014vta} and warped extradimensional models \cite{Hernandez:2015zeh}.
The $Z_{4}$ is the smallest cyclic
symmetry that guarantees that the renormalizable Yukawa terms for the
fermion singlets $S_{i}$ ($i=1,2,3$) only involve the scalar fields
$\xi_{i}$, assumed to acquire very small VEVs.
This feature is crucial to obtain an inverse seesaw contribution to
the light active neutrino mass matrix, instead of a double seesaw
contribution, thus giving rise to heavy quasi-Dirac neutrinos within
the LHC reach.

%Table~\ref{tab:gqn} illustrates the role of various terms in the
%Lagrangean in inducing important fermion mass terms in the model.
%%
%\begin{table}[ht]
%\centering
%\begin{tabular}{|c|c|}
%\hline\hline
%$y_{3}\overline{F}_{3L}\Phi_{1}F_{3R}$ & $\left(M_U\right)_{33}=\left(M_1%
%\right)_{33}$ \\ \hline
%$\kappa_{3}\overline{F}_{3L}\phi_{3}\Psi_{3R}$ & $\left(M_a\right)_{33}$ \\ 
%\hline
%$\gamma_{3}\overline{\Psi }_{3L}\varphi_{3}F_{3R}$ & $\left(M_b\right)_{33}$
%\\ \hline
%$A_i\overline{\Psi }_{L}\Psi_{iR}$, $a_i\overline{\Psi }%
%_{riL}\left( \Sigma \right)_{s}^{r}\Psi_{iR}^{s}\frac{\sigma ^{6}}{\Lambda
%^{6}}$ & $\left(M^{(D)}_c\right)_{ii}$ \ \ \textit{and} \ \ $\left(M^{(l)}_c\right)_{ii}$\\ \hline
%$\frac{\alpha_{3}}{\Lambda }\overline{F}_{L}\chi_{L}\left( S\tau \right)_{%
%\mathbf{1}_{\mathbf{2,0}}}$ & $\left(M_2\right)_{33}$ \\ \hline
%$\frac{\beta_{3}}{\Lambda }F_{R}\chi_{R}^{\dagger }\left( S\tau \right)_{%
%\mathbf{1}_{\mathbf{2,0}}}$ & $\left(M_3\right)_{33}$ \\ \hline
%$\lambda_{1}\left( \overline{S}S^{C}\right)_{\mathbf{3}_{S_{1}}}\xi$ and $%
%\lambda_{2}\left( \overline{S}S^{C}\right)_{\mathbf{3}_{S_{2}}}\xi$ & $\mu$
%\\ \hline\hline
%\end{tabular}%
%\caption{Yukawa terms contributing to some relevant entries of the fermion mass
%matrices.}
%\label{tab:gqn}
%\end{table}

It is worth noting that the Yukawa Lagrangian (\ref{eq:Yukawa-1}) possesses
accidental $U_{1}$-symmetries located in the non-SM sector with the field
charge ($Q$) assignments: 
\begin{eqnarray}
U_{1}^{(a)} &:&Q^{(a)}(S)=1,\ \ Q^{(a)}(\xi )=Q^{(a)}(\zeta )=2,\ \
Q^{(a)}(\rho )=Q^{(a)}(\eta )=Q^{(a)}(\tau )=-1.  \label{eq:U1-assign-1} \\
U_{1}^{(b)} &:&Q^{(b)}(S)=1,\ \ Q^{(b)}(\xi )=Q^{(b)}(\zeta )=2,\ \ \
Q^{(b)}(\chi_{L})=-Q^{(b)}(\chi_{R})=-1.
\end{eqnarray}
These are spontaneously broken by the VEVs of the corresponding scalar
fields in Eq.~(\ref{VEVpattern}). As a result there appear massless
Goldstone bosons with interaction strengths determined by 
the VEVs shown in Eq.~(\ref{VEVpattern}).
This leads to the presence of invisible Higgs
decays~\cite{joshipura:1992hp} which are restricted by LEP as well as
LHC searches~\cite{Bonilla:2015jdf}.
From Eq.~(\ref{VEVhierarchy}) one sees that the $U_{1}^{(a)}$ symmetry
breaking scale is large of the order of $\Lambda_{PS}\sim 10^{6}$ GeV,
while the $U_{1}^{(b)}$ symmetry is broken at the low scale
$v_{\xi }\sim v_{\zeta }$. Thus the $U_{1}^{(b)} $-Goldstone can have
a significant couplings with the fields in the exotic sector and these
interactions could potentially leak via mixing to the SM sector. On
the other hand, as previously mentioned in section \ref{sec:model},
the 125 GeV Higgs boson is dominantly composed of
$\left( \Phi_{1}\right)_{11}$, which is the only scalar contributing
to the top quark mass. Since the $U_{1}^{(b)}$ symmetry breaking scale
$v_{\xi }\sim v_{\zeta }$ is much smaller than the scale of
electroweak symmetry breaking scale $v=246$ GeV, the mixing of the
$U_{1}^{(b)}$-Goldstone with the 125 GeV Higgs boson is suppressed by
the ratios of their VEVs
(c.f. Ref. \cite{Grimus:2000vj}). 
Alternatively, these Goldstones may also be avoided by adding explicit
breaking trilinear terms in the scalar potential. A detailed study is
beyond the scope of the present paper. % and will be addressed elsewhere.

\section{Quark masses and mixings}

\label{quarkmassesandmixing}

From the first line in Eq.~(\ref{eq:Yukawa-1}), the up-type quark mass matrix
is given by
\begin{equation}
M_{U}=\left( 
\begin{array}{ccc}
y_{1}\lambda ^{8} & 0 & 0 \\ 
0 & y_{2}\lambda ^{4} & 0 \\ 
0 & 0 & y_{3}%
\end{array}%
\right) \allowbreak \frac{v}{\sqrt{2}},
\end{equation}
where $y_{i}$ ($i=1,2,3$) are $\mathcal{O}(1)$ parameters and we set $%
v_{1}^{\left( 1\right) }=v_{1}^{\left( 2\right) }=\frac{v}{\sqrt{2}}$, with $%
v=246$ GeV the scale of electroweak symmetry breaking and $v_{\sigma
}=\lambda \Lambda $, with $\lambda =0.225$ being one of the Wolfenstein
parameters.

For the sake of simplicity, we assume $v_{2}^{\left( j\right) }=0$
($j=1,2$) so that the \sm down-type quarks and charged leptons acquire
their masses from a universal seesaw mechanism, mediated by the exotic
down-type quarks $D_i$ and charged leptons $L_i$ present in $\Psi_{iR}$ and
$\overline{\Psi }_{iL}$. In this case the down-type charged fermion
mass matrices take the form
\begin{equation}
\left( 
\begin{array}{cc}
\overline{d}_{iL} &\overline{D}_{iL} 
\end{array}
\right) \left( 
\begin{array}{cc}
0_{3\times 3} & M_{a} \\ 
M_{b} & M_{D}%
\end{array}%
\right) 
 \left( 
\begin{array}{c}
d_{jR}  \\ 
D_{jR}
\end{array}%
\right)\, ,\qquad
\left( 
\begin{array}{cc}
\overline{l}_{iL} &\overline{L}_{iL} 
\end{array}
\right) \left( 
\begin{array}{cc}
0_{3\times 3} & M_{a} \\ 
M_{b} & M_{l}%
\end{array}%
\right) 
 \left( 
\begin{array}{c}
l_{jR}  \\ 
L_{jR}
\end{array}%
\right),
\end{equation}
with
\begin{eqnarray}
M_{a}&=&\left( 
\begin{array}{ccc}
\kappa_{1}\lambda ^{2} & 0 & 0 \\ 
0 & \kappa_{2}\lambda & 0 \\ 
0 & 0 & \kappa_{3}%
\end{array}%
\right) \allowbreak v_{\phi }\,,\qquad  M_{b}=\left( 
\begin{array}{ccc}
\gamma_{1}\lambda ^{2} & \gamma_{4}\lambda & \gamma_{5}\lambda \\ 
0 & \gamma_{2}\lambda & \gamma_{6}\lambda \\ 
0 & 0 & \gamma_{3}%
\end{array}%
\right) \allowbreak v_{\varphi }\,,\nonumber\\
M_{D} &=&\left( 
\begin{array}{ccc}
A_{1}+\frac{v_{\Sigma_{2}}}{2\sqrt{6}}a_{1} & 0 & 0 \\ 
0 & A_{2}+\frac{v_{\Sigma_{2}}}{2\sqrt{6}}a_{2} & 0 \\ 
0 & 0 & A_{3}+\frac{v_{\Sigma_{2}}}{2\sqrt{6}}a_{3}%
\end{array}%
\right)\,, \nonumber\\ M_{l}&=&\left( 
\begin{array}{ccc}
A_{1}-\frac{3v_{\Sigma_{2}}}{2\sqrt{6}}a_{1} & 0 & 0 \\ 
0 & A_{2}-\frac{3v_{\Sigma_{2}}}{2\sqrt{6}}a_{2} & 0 \\ 
0 & 0 & A_{3}-\frac{3v_{\Sigma_{2}}}{2\sqrt{6}}a_{3}%
\end{array}%
\right)\,,
\end{eqnarray}
and the further simplification $v_{\phi_{i}}=v_{\phi }$ and $v_{\varphi
_{l}}=v_{\varphi }$.

Taking the limit $ M_{a}$, $ M_{b} \ll
A_{i}$, the \sm down-type quark and charged lepton mass matrices become
\begin{equation}  \label{MDl}
M_{f}^{\left( 1\right) }=M_{a}\left( M_{f}\right)
^{-1}M_{b}=\left( 
\begin{array}{ccc}
a_{1f}\lambda ^{7} & a_{4f}\lambda ^{6} & a_{5f}\lambda ^{6} \\ 
0 & a_{2f}\lambda ^{5} & a_{6f}\lambda ^{5} \\ 
0 & 0 & a_{3f}\lambda ^{3}%
\end{array}%
\right) \frac{v}{\sqrt{2}},\hspace{1cm}f=D,l.
\end{equation}
where we have set
$v_{\phi }=\lambda ^{3}\frac{v}{\sqrt{2}}\frac{m_{\Psi }}{%
  v_{\varphi }}$.

Notice that in our model the CKM matrix arises only from the down-type
quark sector. In order to recover the low energy quark flavor data, we
assume that all dimensionless parameters of the SM down-type quark
mass matrix are real, excepting $a_{5D}$, taken to be complex. 
The physical quark mass spectrum \cite{Bora:2012tx,Xing:2007fb} and mixing angles \cite{Olive:2016xmw} can be perfectly reproduced
in terms of natural parameters of order one, as shown in Table
\ref{Tab}, starting from the following benchmark point:
\begin{equation}
\begin{array}{c}
y_{1}\simeq 1.269\,,\hspace{1cm}y_{2}\simeq 1.424\,,\hspace{1cm}y_{3}\simeq
0.989\,, \\ 
a_{1D}\simeq 0.585\,,\hspace{1cm}a_{2D}\simeq 0.560\,,\hspace{1cm}%
a_{3D}\simeq 1.421\,, \\ 
a_{4D}\simeq 0.573\,,\hspace{1cm}\left\vert a_{5D}\right\vert \simeq 0.446\,,%
\hspace{1cm}\arg (a_{5D})\simeq -1.906\,,\hspace{1cm}a_{6D}\simeq 1.153\,.%
\end{array}
\label{eq:bm-values}
\end{equation}
\begin{table}[tbh]
\begin{center}
\begin{tabular}{c|l|l}
\hline\hline
Observable & Model value & Experimental value \\ \hline
$m_{u}(\mathrm{MeV})$ & \quad $1.45$ & \quad $1.45_{-0.45}^{+0.56}$ \\ \hline
$m_{c}(\mathrm{MeV})$ & \quad $635$ & \quad $635\pm 86$ \\ \hline
$m_{t}(\mathrm{GeV})$ & \quad $172$ & \quad $172.1\pm 0.6\pm 0.9$ \\ \hline
$m_{d}(\mathrm{MeV})$ & \quad $2.9$ & \quad $2.9_{-0.4}^{+0.5}$ \\ \hline
$m_{s}(\mathrm{MeV})$ & \quad $57.7$ & \quad $57.7_{-15.7}^{+16.8}$ \\ \hline
$m_{b}(\mathrm{GeV})$ & \quad $2.82$ & \quad $2.82_{-0.04}^{+0.09}$ \\ \hline
$\sin \theta_{12}$ & \quad $0.225$ & \quad $0.225$ \\ \hline
$\sin \theta_{23}$ & \quad $0.0411$ & \quad $0.0411$ \\ \hline
$\sin \theta_{13}$ & \quad $0.00357$ & \quad $0.00357$ \\ \hline
$\delta $ & \quad $1.236$ & \quad $1.236$ \\ \hline\hline
\end{tabular}%
\end{center}
\caption{Model and experimental values of the quark masses and CKM
parameters.}
\label{Tab}
\end{table}

To close this section we briefly comment on the phenomenological
implications of our model regarding flavor changing processes
involving quarks.  It is noteworthy that the flavor changing top quark
decays $t \to hc$ and $t \to hu$ are absent at tree level in our
model, as follows from the quark Yukawa terms. The flavor changing top
quark decays $t \to hc$ and $t \to hu$ are induced at one loop level
from virtual charged Higgses and SM down-type quarks running in the
internal lines of the loops. Consequently, a measurement of the
branching fraction for the $t \to hc$ and $t \to hu$ decays at the LHC
may test our model. 
On the other hand the admixture of exotic down-type quarks in the mass
matrix implies a violation of the Glashow-Iliopoulos-Maiani mechanism
in this sector. This may be relevant in connection with the recent B
anomalies~\cite{Bifani:2017gyn}.
%
%This is beyond the scope of this work and is left for future studies.
%
% It would also be interesting to perform a study of the charged Higgs
% production at the LHC, the charged scalars decay modes, and the flavor
% changing top quark decays.

\section{Lepton masses, mixing  and oscillations}
\label{leptonmassesandmixing} 

Here is where the predictive power of our flavor symmetry model is
mainly manifest. From the neutrino Yukawa terms, we obtain the following
neutrino mass terms :
\begin{equation}
-\mathcal{L}_{mass}^{\left( \nu \right) }=\frac{1}{2}\left( 
\begin{array}{ccc}
\overline{\nu_{L}^{C}} & \overline{\nu_{R}} & \overline{S}%
\end{array}%
\right) M_{\nu }\left( 
\begin{array}{c}
\nu_{L} \\ 
\nu_{R}^{C} \\ 
S^{C}%
\end{array}%
\right) +H.c,  \label{Lnu}
\end{equation}
where the neutrino mass matrix reads 
\begin{eqnarray}
M_{\nu } &=&\left( 
\begin{array}{ccc}
0_{3\times 3} & M_{1} & M_{2} \\ 
M_{1}^{T} & 0_{3\times 3} & M_{3} \\ 
M_{2}^{T} & M_{3}^{T} & \mu 
\end{array}%
\right) =\left( 
\begin{array}{ccc}
0_{3\times 3} & M_{U} & \frac{\sqrt{2}v_{L}v_{\rho }}{\Lambda v}M_{U} \\ 
M_{U}^{T} & 0_{3\times 3} & \frac{\sqrt{2}v_{R}v_{\rho }}{\Lambda v}M_{U} \\ 
\frac{\sqrt{2}v_{L}v_{\rho }}{\Lambda v}M_{U}^{T} & \frac{\sqrt{2}%
v_{R}v_{\rho }}{\Lambda v}M_{U}^{T} & \mu 
\end{array}%
\right) \nonumber\\&=&\left( 
\begin{array}{ccc}
0_{3\times 3} & M_{U} & \frac{v_{L}}{v_{R}}M_{U} \\ 
M_{U}^{T} & 0_{3\times 3} & M_{U} \\ 
\frac{v_{L}}{v_{R}}M_{U}^{T} & M_{U}^{T} & \mu 
\end{array}%
\right) ,\hspace{1cm}  
\end{eqnarray}
with
\begin{equation}
\mu  =\left( 
\begin{array}{ccc}
r\lambda_{1} & \lambda_{2}e^{i\psi } & \lambda_{2}e^{-i\psi } \\ 
&  &  \\ 
\lambda_{2}e^{i\psi } & \lambda_{1}e^{-i\psi } & r\lambda_{2} \\ 
&  &  \\ 
\lambda_{2}e^{-i\psi } & r\lambda_{2} & \lambda_{1}e^{i\psi }%
\end{array}%
\right) \allowbreak \frac{v_{\xi }}{\sqrt{2+r^{2}}}\,,
\end{equation}
where for the sake of simplicity, we set $\alpha_{i}=\beta_{i}=y_{i}$ ($%
i=1,2,3$) and $v_{\rho }=v_{\eta }=v_{\tau }=\frac{v}{\sqrt{2}v_{R}}\Lambda $.
Then, 
\begin{equation}
M_{\nu }^{\left( 1\right) }=\mu -\frac{v_{L}}{v_{R}}\left(
M_{U}+M_{U}^{T}\right) ,\quad M_{\nu }^{\left( 2\right) }=-\frac{1}{2}%
\left( M_{U}+M_{U}^{T}\right) +\frac{1}{2}\mu ,\quad M_{\nu }^{\left(
3\right) }=\frac{1}{2}\left( M_{U}+M_{U}^{T}\right) +\frac{1}{2}\mu
\label{Mnu1}
\end{equation}
where $M_{\nu }^{\left( 1\right) }$ corresponds to the physical light
neutrino mass matrix whereas $M_{\nu }^{\left( 2\right) }$ and
$M_{\nu }^{\left( 3\right) } $ are the heavy quasi-Dirac neutrino mass
entries. Note that the physical eigenstates include three active
neutrinos and six heavy, mainly isosinglet, neutrinos. The heavy
quasi-Dirac neutrinos have a small splitting $\mu $.

In the limit of vanishing contributions from linear seesaw
($v_L \ll v_R$), the light neutrino mass matrix becomes
\begin{equation}
M_{\nu }^{\left( 1\right) }\approx \mu =\left( 
\begin{array}{ccc}
r \lambda_{1} & \lambda_{2}e^{i\psi } & \lambda_{2}e^{-i\psi } \\ 
&  &  \\ 
\lambda_{2}e^{i\psi } & \lambda_{1}e^{-i\psi } & r \lambda_{2} \\ 
&  &  \\ 
\lambda_{2}e^{-i\psi } & r \lambda_{2} & \lambda_{1}e^{i\psi }%
\end{array}%
\right) \allowbreak \frac{v_{\xi }}{\sqrt{2+r^2}}.
\end{equation}
Taking real Yukawa couplings $\lambda_i$ and VEVs,
$M_{\nu }^{\left( 1\right) }$ display explicit generalized $\mu-\tau$
symmetry~\cite{babu:2002dz,grimus:2003yn,King:2014nza}
\begin{equation}
X^{T}M_{\nu }^{\left( 1\right) } X =M_{\nu }^{\left( 1\right)* },
\end{equation}
with 
\begin{equation}
X=\left( 
\begin{array}{ccc}
1 & 0 & 0 \\ 
0 & 0 & 1 \\ 
0 & 1 & 0%
\end{array}
\right).
\end{equation}
The most general matrix $V_{\nu}$ that diagonalizes $M_{\nu }^{\left(
1\right) }$ through $V_{\nu}^{T}M_{\nu }^{\left( 1\right) }V_{\nu}=\mathrm{%
diag}(m^{\nu}_{1},m^{\nu}_{2},m^{\nu}_{3})$ is given by \cite%
{Chen:2015siy,Chen:2016ica} 
\begin{equation}
V_{\nu}=\Sigma O_{23} O_{13} O_{12}Q_{\nu},
\end{equation}
where 
\begin{equation}
\Sigma=\left( 
\begin{array}{ccc}
1 & 0 & 0 \\ 
0 & \frac{1}{\sqrt{2}} & \frac{i}{\sqrt{2}} \\ 
0 & \frac{1}{\sqrt{2}} & -\frac{i}{\sqrt{2}}%
\end{array}
\right),
\end{equation}
is the Takagi factorization matrix of $X$ defined as $X=\Sigma \Sigma^{T}$ , 
$O_{ij}$ are $3\times 3$ orthogonal matrix parameterized as 
\begin{equation}  \label{Ort}
\begin{split}
O_{23}&=\left( 
\begin{array}{ccc}
1 & 0 & 0 \\ 
0 & \cos \omega_{23} & \sin \omega_{23} \\ 
0 & -\sin \omega_{23} & \cos \omega_{23}%
\end{array}
\right)\,,\qquad O_{13}= \left( 
\begin{array}{ccc}
\cos \omega_{13} & 0 & \sin \omega_{13} \\ 
0 & 1 & 0 \\ 
-\sin \omega_{13} & 0 & \cos \omega_{13}%
\end{array}
\right)\,,\\\qquad O_{12}&= \left( 
\begin{array}{ccc}
\cos \omega_{12} & \sin \omega_{12} & 0 \\ 
-\sin \omega_{12} & \cos \omega_{12} & 0 \\ 
0 & 0 & 1%
\end{array}
\right),
\end{split}
\end{equation}
and $Q_{\nu}=\mathrm{diag}(e^{- i \pi k_1/2},e^{- i \pi k_2/2},e^{- i \pi
k_3/2})$ is a diagonal matrix with $k_i=0,1,2,3$.

Notice that, with real Yukawa couplings $\lambda_i$ and VEVs in
Eq.(\ref{Mnu1}), we have a reduced number of parameters, leading to
the following relations:
\begin{equation}  \label{pr}
\begin{split}
&\tan2\omega_{12}=\frac{4 \sin \omega_{13}\left[\sin \left(2 \psi +\omega
_{23}\right)+\sin 3 \omega_{23}\right]}{\left(3 \cos 2 \omega
_{13}-1\right) \cos \left(2 \psi +\omega_{23}\right)+\left(\cos 2 \omega
_{13}-3\right) \cos 3 \omega_{23}}\,, \\
&\lambda_1=-\sqrt{2} \lambda_2 \tan \omega_{13} \cos \left(\psi -\omega
_{23}\right) \csc \left(\psi +2 \omega_{23}\right) \\
&r=-\frac{\sqrt{2} \left[\sin \omega_{13} \cos \left(\psi -\omega
_{23}\right) \cos \left(\psi +2 \omega_{23}\right)+2 \cos \omega_{13} \cot
2 \omega_{13} \sin \left(\psi -\omega_{23}\right) \sin \left(\psi +2
\omega_{23}\right)\right]}{\sqrt{2} \sin \omega_{13} \cos \left(\psi
-\omega_{23}\right)+\cos \omega_{13} \sin \left(\psi +2 \omega_{23}\right)%
}\,.
\end{split}%
\end{equation}

On the other hand, the light charged lepton mass matrix in
Eq. (\ref{MDl}) is diagonalized by an almost diagonal unitary matrix
through
$V_{l}^{\dagger}M^{\left( 1\right) }_{l}M_{l}^{\left( 1\right)
  \dagger} V_{l}=%
\mathrm{diag}(m^2_{e},m^2_{\mu},m^2_{\tau})$. Taking real entries in
$M^{\left( 1\right)}_{l}$, at first approximation $V_{l}$ is dominated
by the Cabibbo angle and can be written as
\begin{equation}
V_{l}\approx\left( 
\begin{array}{ccc}
\cos \theta_0 & \sin \theta_0 & 0 \\ 
-\sin \theta_0 & \cos \theta_0 & 0 \\ 
0 & 0 & 1%
\end{array}
\right) ,\qquad \sin \theta_0\approx\lambda.
\end{equation}

In this approximation, the lepton mixing matrix is given as
\begin{equation}
U=V_{l}^{\dagger}V_{\nu}.
\end{equation}
In the fully ``symmetrical'' presentation of the lepton mixing
matrix~\cite{Schechter:1980gr,Rodejohann:2011vc}
\begin{equation}  \label{eq:symmetric_para}
U = \left( 
\begin{array}{ccc}
c_{12} c_{13} & s_{12} c_{13} e^{ - i \phi_{12} } & s_{13} e^{ -i \phi_{13} }
\\ 
-s_{12} c_{23} e^{ i \phi_{12} } - c_{12} s_{13} s_{23} e^{ -i ( \phi_{23} -
\phi_{13} ) } & c_{12} c_{23} - s_{12} s_{13} s_{23} e^{ -i ( \phi_{23} +
\phi_{12} - \phi_{13} ) } & c_{13} s_{23} e^{- i \phi_{23} } \\ 
s_{12} s_{23} e^{ i ( \phi_{23} + \phi_{12} ) } - c_{12} s_{13} c_{23} e^{ i
\phi_{13} } & - c_{12} s_{23} e^{ i \phi_{23} } - s_{12} s_{13} c_{23} e^{
-i ( \phi_{12} - \phi_{13} ) } & c_{13} c_{23}%
\end{array}
\right)\,,
\end{equation}
with $c_{ij}=\cos\theta_{ij}$ and $s_{ij}=\sin\theta_{ij}$, the relation
between flavor mixing angles and the magnitudes of the entries of the
leptonic mixing matrix is 
\begin{equation}  \label{eq:UU}
\sin^{2} \theta_{13} = \left| U_{e3} \right|^{2} \, , \quad \sin^{2}
\theta_{12} = \frac{ \left| U_{e2} \right|^{2} }{ 1 - \left| U_{e3}
\right|^{2} } \quad \text{and} \quad \sin^{2} \theta_{23} = \frac{ \left|
U_{\mu 3} \right|^{2} }{ 1 - \left| U_{e3} \right|^{2} } \,.
\end{equation}
The Jarlskog invariant is defined as $J_{\mathrm{CP}}=\mathrm{Im}\left(
U_{11}^{*} U_{23}^{*} U_{13} U_{21} \right)$, %%
and takes the form~\cite{Rodejohann:2011vc}
\begin{equation}
J_{\mathrm{CP}} = \frac{1}{8} \sin 2 \theta_{12} \, \sin 2 \theta_{23} \,
\sin 2 \theta_{13}\, \cos\theta_{13} \,\sin(\phi_{13}-\phi_{23}-\phi_{12})
\,,
\end{equation}
whereas the two additional Majorana-type rephasing invariants
$I_{1} = \mathrm{Im}\left( U_{12}^{2} U_{11}^{* 2} \right)$ and
$I_{2} = \mathrm{Im}\left( U_{13}^{2} U_{11}^{*2} \right)$, become
\begin{equation}
\begin{array}{l}
I_{1} = \frac{1}{4} \sin^{2} 2\theta_{12} \cos^{4} \theta_{13} \sin ( - 2
\phi_{12} ) \quad \text{and} \quad I_{2} = \frac{1}{4} \sin^{2} 2
\theta_{13} \cos^{2} \theta_{12} \sin ( - 2 \phi_{13} )\,.%
\end{array}%
\end{equation}

In terms of the model parameters, the lepton mixing angles are expressed as 
\begin{equation}  \label{s13}
\sin^2 \theta_{13}=\frac{1}{4} \left(4 \cos ^2\theta_0 \sin ^2\omega
_{13}+2 \sin ^2\theta_0 \cos ^2\omega_{13}-\sqrt{2} \sin 2 \theta_0 \sin
2 \omega_{13} \sin \omega_{23}\right)\,,
\end{equation}
\begin{equation}  \label{s12}
\begin{split}
\sin^2 \theta_{12}=&\big[2 \sin ^2\theta_0 \left(\sin ^2\omega_{12} \sin
^2\omega_{13}+\cos ^2\omega_{12}\right)+4 \cos ^2\theta_0 \sin ^2\omega
_{12} \cos ^2\omega_{13} \\
&+\sqrt{2} \sin 2 \theta_0 \left(\sin ^2\omega_{12} \sin 2 \omega_{13}
\sin \omega_{23}-\sin 2 \omega_{12} \cos \omega_{13} \cos \omega
_{23}\right)\big] \\
& \times \big[4-4 \cos ^2\theta_0 \sin ^2\omega_{13}-2 \sin ^2\theta_0
\cos ^2\omega_{13}+\sqrt{2} \sin 2 \theta_0 \sin 2 \omega_{13} \sin
\omega_{23}\big]^{-1}\,,
\end{split}%
\end{equation}
\begin{equation}  \label{s23}
\sin^2 \theta_{23}=1-\frac{2 \cos ^2\omega_{13}}{4-4 \cos ^2\theta_0 \sin
^2\omega_{13}-2 \sin ^2\theta_0 \cos ^2\omega_{13}+\sqrt{2} \sin 2 \theta
_0 \sin 2 \omega_{13} \sin \omega_{23}} \,,
\end{equation}
and our predicted correlations: 
\begin{equation}  \label{corr1}
\cos ^2\theta_{13} \cos ^2\theta_{23}=\frac{1}{2} \cos ^2\omega_{13}\,,
\end{equation}
\begin{equation}  \label{corr2}
\cos 2 \theta_{12} \cos ^2\theta_{13}=\cos 2 \omega_{12} \left(\cos
^2\theta_{13}-\sin ^2\theta_0\right)+\frac{1}{\sqrt{2}}\sin 2 \theta_0
\sin 2 \omega_{12} \cos \omega_{13} \cos \omega_{23}\,.
\end{equation}
On the other hand, the rephasing invariant CP violation parameter combinations are 
\begin{equation}  \label{JCP}
\begin{split}
J_{\mathrm{CP}}=&\frac{1}{64} \bigg\{-\sqrt{2} \sin 2 \theta_0 \left[4 \sin
2 \omega_{13} \cos 2 \omega_{12} \cos \omega_{23}+\sin 2 \omega_{12}
\sin \omega_{23} \left(\cos \omega_{13}+3 \cos 3 \omega_{13}\right)\right]
\\
&\qquad+16 \cos 2 \theta_0 \sin 2 \omega_{12} \sin \omega_{13} \cos
^2\omega_{13}\bigg\}.
\end{split}%
\end{equation}
\begin{equation}
\begin{split}
  I_{1}=&\frac{(-1)^{k_1+k_2}}{8} \sin \theta_0 \cos \omega_{13}
  \bigg\{4 \sqrt{2} \cos ^3\theta_0 \sin 2 \omega_{12} \sin \omega
 _{23} \cos^2\omega_{13}-\sin^3\theta_0 \sin 2 \omega_{12} \sin 2 \omega_{13} \\
  &+2 \sin \theta_0 \cos ^2\theta_0 \left[\sin 2 \omega_{12} \sin 2 \omega
   _{13} \left(2-\cos 2 \omega_{23}\right)-2 \sin 2 \omega_{23} \cos 2 \omega_{12} \cos \omega_{13}\right] \\
  &+\sqrt{2} \sin ^2\theta_0 \cos \theta_0 \left[\sin 2 \omega_{12}
    \sin \omega_{23} \left(1-3 \cos 2 \omega_{13}\right)-4 \sin
    \omega_{13} \cos 2 \omega_{12} \cos \omega_{23}\right]\bigg\}
\end{split}
\end{equation}
\begin{equation}
\begin{split}
I_{2}=& \frac{(-1)^{k_1+k_3}}{32} \bigg\{\sin 2 \theta_0 \big\{ \sqrt{2}%
\sin 2 \omega_{12} \sin \omega_{23} \cos \omega_{13} \left[\cos 2 \theta
_0 \left(3-5 \cos 2 \omega_{13}\right)+2 \cos ^2\omega_{13}\right] \\
&-4\sqrt{2} \sin 2 \omega_{13} \cos \omega_{23} \left(\cos 2 \theta_0
\cos 2 \omega_{12}+\cos ^2\theta_0\right)\\&+\sin 2 \theta_0 \sin 2 \omega
_{23} \left[2 \sin ^2\omega_{12}+\left(3 \cos 2 \omega_{12}+1\right) \cos
2 \omega_{13}\right]\big\} \\
&-4 \sin ^22 \theta_0 \sin 2 \omega_{12} \sin ^3\omega_{13} \cos 2 \omega
_{23}-4 \sin ^2\theta_0 \left(5 \cos 2 \theta_0+3\right) \sin 2 \omega
_{12} \sin\omega_{13} \cos ^2\omega_{13}\bigg\}\,.
\end{split}%
\end{equation}
Eliminating $\omega_{12}$ in the above relations using eq.(\ref{pr}),
the mixing parameters depend ultimately on three angles $\omega_{23}$,
$\omega _{13}$, $\psi$ up to three discrete variables $k_1$, $k_2$,
$k_3$.  Furthermore, without loss of generality these angles are
restricted to $\omega_{23}\in[-\pi,\pi]$,
$\omega_{13}\in[-\pi/2,\pi/2]$ and $\psi\in[-\pi/2,\pi/2]$.
Notice that the angle $\psi$ in this framework is responsible for the CP
violating phase in the lepton sector, since the first stage of the
diagonalization process yields a real symmetric matrix
\begin{equation}
\Sigma^{T}M_{\nu }^{\left( 1\right) }\Sigma=\frac{v_{\xi }}{\sqrt{2+r^2}}%
\left( 
\begin{array}{ccc}
r\lambda_1 & \sqrt{2}\lambda_2 \cos \psi & -\sqrt{2}\lambda_2 \sin \psi
\\ 
\sqrt{2}\lambda_2 \cos \psi & r\lambda_2+\lambda_1\cos \psi & \lambda
_1\sin \psi \\ 
-\sqrt{2}\lambda_2 \sin \psi & \lambda_1\sin \psi & r\lambda_2-\lambda
_1\cos \psi \\ 
&  & 
\end{array}
\right)\,,
\end{equation}
and in the limit of vanishing $\psi$, this matrix is already block
diagonal, implying $\omega_{23}=\omega_{13}=0$ in Eq.(\ref{JCP}),
which in turn leads to $J_{\mathrm{CP}}=0$. 
Moreover, from Eq.(\ref{corr1}), notice that if $\sin^2 \theta_{13}$
is allowed to vary within $3\sigma$ according to the global fit
\cite{Forero:2014bxa}, then $\sin^2 \theta_{23}$ is restricted to the
range $(0.487, 0.539)$.\\

\begin{figure}[!ht]
\centering
\includegraphics[width=0.45\textwidth]{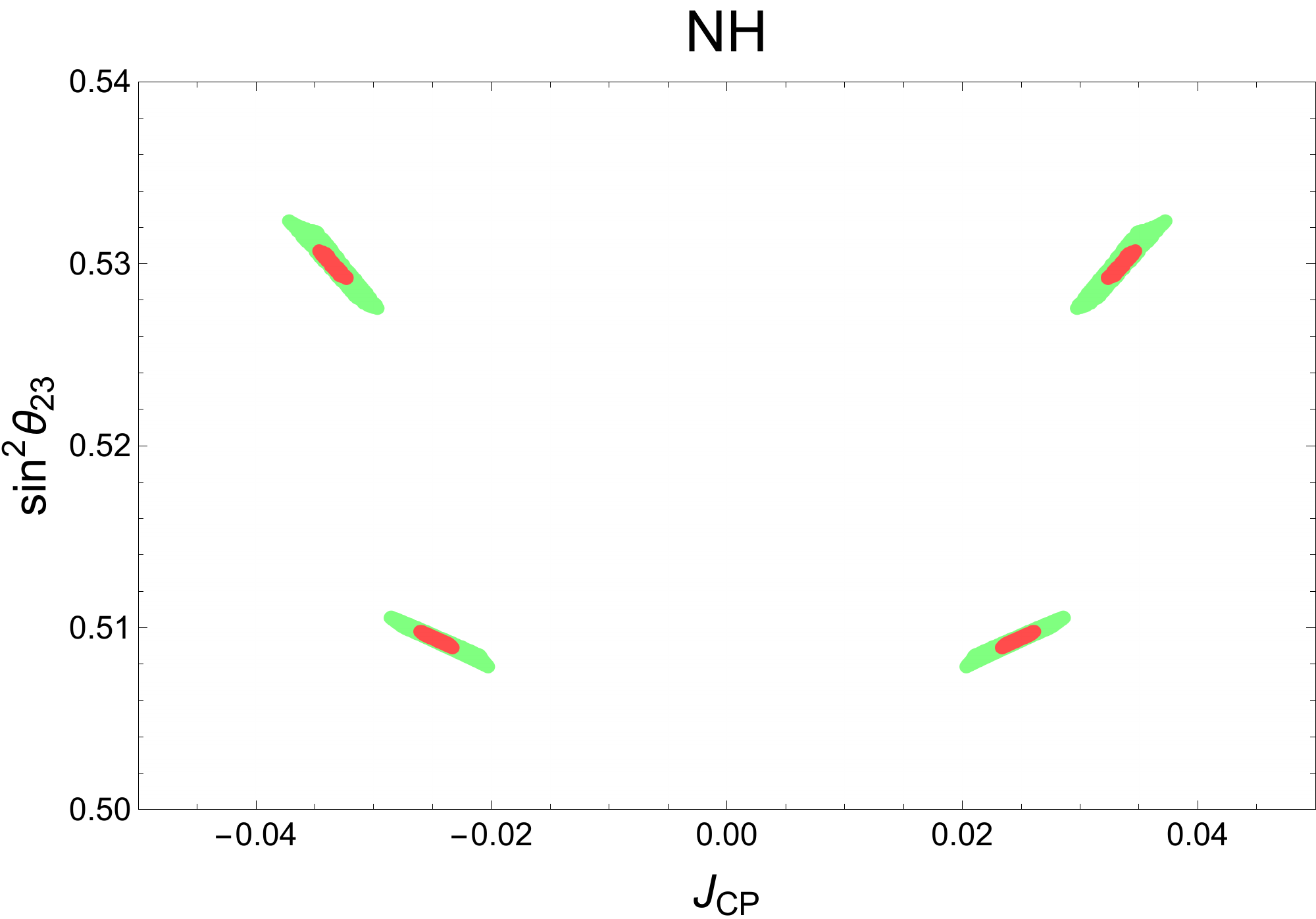}
\caption{Predicted CP violation versus atmospheric angle for NH. 
  Green (Red) regions correspond to 3 (1)
  $\protect\sigma$ values for the solar and reactor angles in
  Eqs.~(\protect\ref{s13}), (\protect\ref{s12}). All predicted values for
  $\sin^2 \protect\theta_{23}$ lie inside its 1$\protect\sigma$
  region, according to the global fit \protect\cite{Forero:2014bxa}. }
\label{NHplot}
\end{figure}
\begin{figure}[!ht]
\centering
\includegraphics[width=0.45\textwidth]{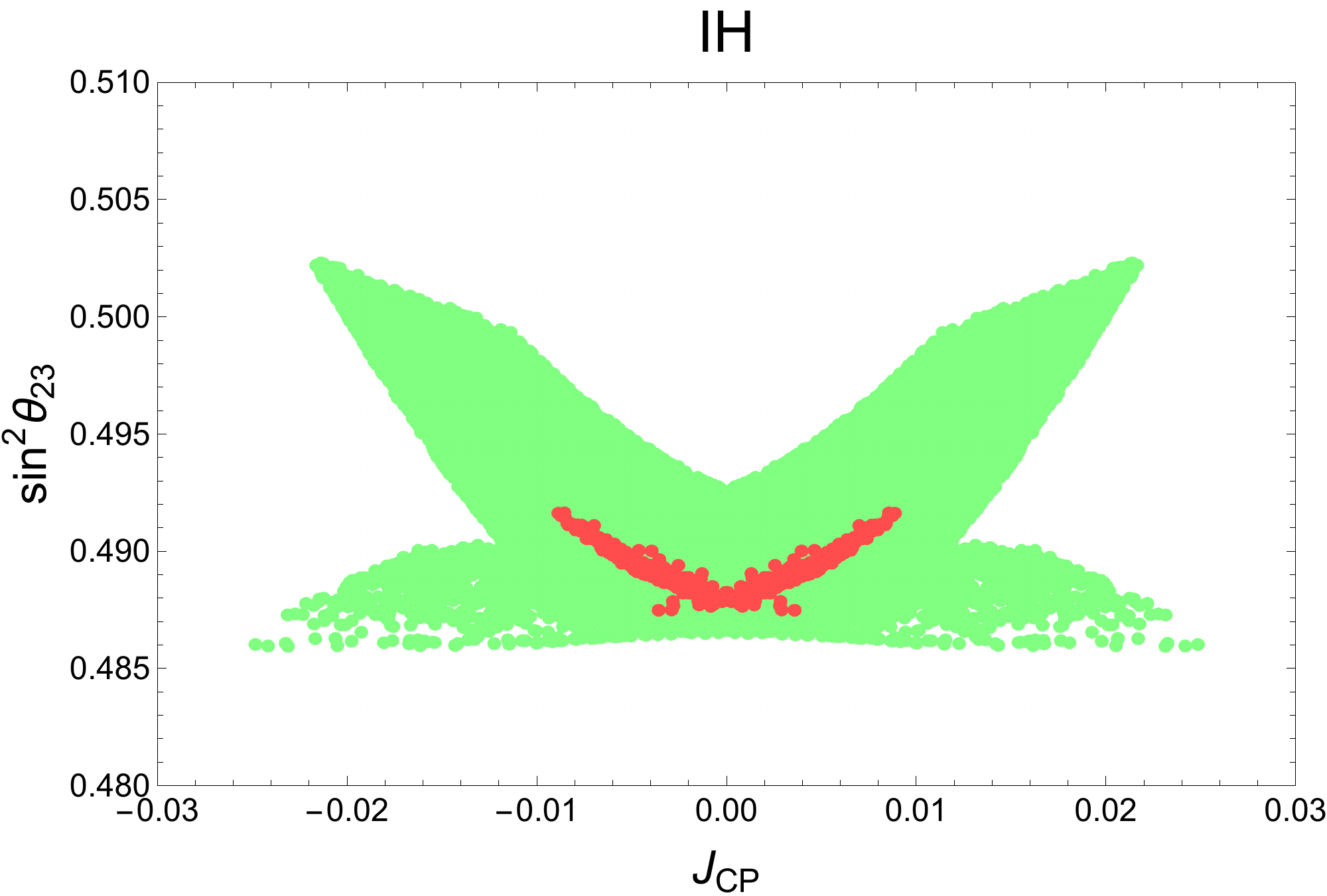}
\caption{Predicted CP violation versus atmospheric angle for IH. Green
  (Red) regions correspond to 3 (1) $\protect\sigma$ values for the
  solar and reactor angles in Eqs.~(\ref{s13}), (\ref{s12}). All predicted
  values for $\sin^2 \protect\theta_{23}$ in this case lie into its
  2$\sigma$ region \protect\cite{Forero:2014bxa}.}
\label{IHplot}
\end{figure}

In figures \ref{NHplot} and \ref{IHplot}, we give the allowed values
for $\sin^2 \theta_{23}$ together with the corresponding
$J_{\text{CP}}$ predictions in both mass orderings.
For our analysis, we randomly generated parameter configurations
for $\omega_{23}$, $\omega_{13}$ and $\psi$ corresponding to
3(1)$\protect\sigma$ values for the solar and reactor angles in
Eqs.~(\protect\ref{s13}),(\protect\ref{s12}).
One sees that for the Normal Hierarchy (NH) case the allowed region is
severely restricted, with a fourfold degeneracy. In this case CP must
necessarily be violated in oscillations, and the predicted atmospheric
angle lies in the higher octant, inside its $1\sigma$ region.

In contrast, for the case of Inverted Hierarchy (IH) one sees that CP
can be conserved in neutrino oscillations. Moreover, if violated, it
is unlikely for CP to be maximally violated. The predicted atmospheric
angle lies inside its $2\sigma$ region, preferably in the first
octant.

\section{Neutrinoless double beta decay}
\label{sec:neutr-double-beta}

In this section we determine the effective Majorana neutrino
mass parameter characterizing the neutrinoless double beta
($0\nu\beta\beta$)
decay amplitude. It is given by:
\begin{equation}
\begin{split}
\left\vert m_{\beta\beta}\right\vert =& \left\vert
\sum_{i=1}^{3}m^{\nu}_{i}U_{ei}^{2}\right\vert \\
=& \frac{1}{4}\bigg|m_{1}\left( 2\cos \theta_{0}\cos \omega_{12}\cos
\omega_{13}+\sqrt{2}e^{-i\omega_{23}}\sin \theta_{0}\left[ \sin \omega
_{12}+i\sin \omega_{13}\cos \omega_{12}\right] \right) {}^{2} \\
& + m_{2}\left( -2\cos \theta_{0}\sin \omega_{12}\cos \omega_{13}+\sqrt{2}%
e^{-i\omega_{23}}\sin \theta_{0}\left[ \cos \omega_{12}-i\sin \omega
_{12}\sin \omega_{13}\right] \right) {}^{2} \\
& - m_{3}\left( \sqrt{2}e^{-i\omega_{23}}\sin \theta_{0}\cos \omega
_{13}+2i\cos \theta_{0} \sin \omega_{13}\right) {}^{2}\bigg|,
\end{split}
\end{equation}
where $U_{ei}^{2}$ and $m_{\nu_{i}}$ are the lepton mixing matrix
elements and the light active neutrino masses, respectively. The light
active neutrino masses can be written in terms of the parameters of
the model as 
\begin{equation}
\begin{split}
m_{1}=& \frac{\lambda_{2}v_{\xi }}{\sqrt{2+r^{2}}} \bigg\{\frac{\sqrt{2}%
\tan \omega_{13}\sin 2\left( \psi -\omega_{23}\right) +\left( 1-4\tan
^{2}\omega_{13}\right) \cos \left( 2\psi +\omega_{23}\right) -\cos 3\omega
_{23}}{2\tan \omega_{13}\left[ \sqrt{2}\sin \left( \psi +2\omega
_{23}\right) +2\tan \omega_{13}\cos \left( \psi -\omega_{23}\right) \right]
} \\
& \qquad -\frac{\left( 3\cos 2\omega_{13}-1\right) \cos \left( 2\psi
+\omega_{23}\right) +\left( \cos 2\omega_{13}-3\right) \cos 3\omega_{23}}{%
2\sqrt{2}\sin 2\omega_{13}\sin \left( \psi +2\omega_{23}\right) } \\
& \qquad \times \sqrt{1+\frac{16\sin ^{2}\omega_{13}\left[ \sin \left(
2\psi +\omega_{23}\right) +\sin \left( 3\omega_{23}\right) \right] {}^{2}}{%
\left[ \left( 3\cos 2\omega_{13}-1\right) \cos \left( 2\psi +\omega
_{23}\right) +\left( \cos 2\omega_{13}-3\right) \cos 3\omega_{23}\right]
{}^{2}}}\bigg\}\,, \\
m_{2}=& \frac{\lambda_{2}v_{\xi }}{\sqrt{2+r^{2}}} \bigg\{\frac{\sqrt{2}%
\tan \omega_{13}\sin 2\left( \psi -\omega_{23}\right) +\left( 1-4\tan
^{2}\omega_{13}\right) \cos \left( 2\psi +\omega_{23}\right) -\cos 3\omega
_{23}}{2\tan \omega_{13}\left[ \sqrt{2}\sin \left( \psi +2\omega
_{23}\right) +2\tan \omega_{13}\cos \left( \psi -\omega_{23}\right) \right]
} \\
& \qquad +\frac{\left( 3\cos 2\omega_{13}-1\right) \cos \left( 2\psi
+\omega_{23}\right) +\left( \cos 2\omega_{13}-3\right) \cos 3\omega_{23}}{%
2\sqrt{2}\sin 2\omega_{13}\sin \left( \psi +2\omega_{23}\right) } \\
& \qquad \times \sqrt{1+\frac{16\sin ^{2}\omega_{13}\left[ \sin \left(
2\psi +\omega_{23}\right) +\sin \left( 3\omega_{23}\right) \right] {}^{2}}{%
\left[ \left( 3\cos 2\omega_{13}-1\right) \cos \left( 2\psi +\omega
_{23}\right) +\left( \cos 2\omega_{13}-3\right) \cos 3\omega_{23}\right]
{}^{2}}}\bigg\}\,, \\
m_{3}=& \frac{\lambda_{2}v_{\xi }}{\sqrt{2+r^{2}}} \bigg\{\frac{\tan
^{2}\omega_{13}\left[ \cos \left( \psi +2\omega_{23}\right) +\cos 3\psi %
\right] \csc ^{2}\left( \psi +2\omega_{23}\right) -\sqrt{2}\cot \omega
_{13}\sin \left( \psi -\omega_{23}\right) }{1+\sqrt{2}\tan \omega_{13}\cos
\left( \psi -\omega_{23}\right) \csc \left( \psi +2\omega_{23}\right) }%
\bigg\}\,,
\end{split}
\end{equation}
with
$$Q_{\nu}^{T}\mathrm{diag}(m_{1},m_{2},m_{3})Q_{\nu}=\mathrm{diag}
(m^{\nu}_{1},m^{\nu}_{2},m^{\nu}_{3}).$$

\begin{figure}[ht]
\centering
\includegraphics[width=0.45\textwidth]{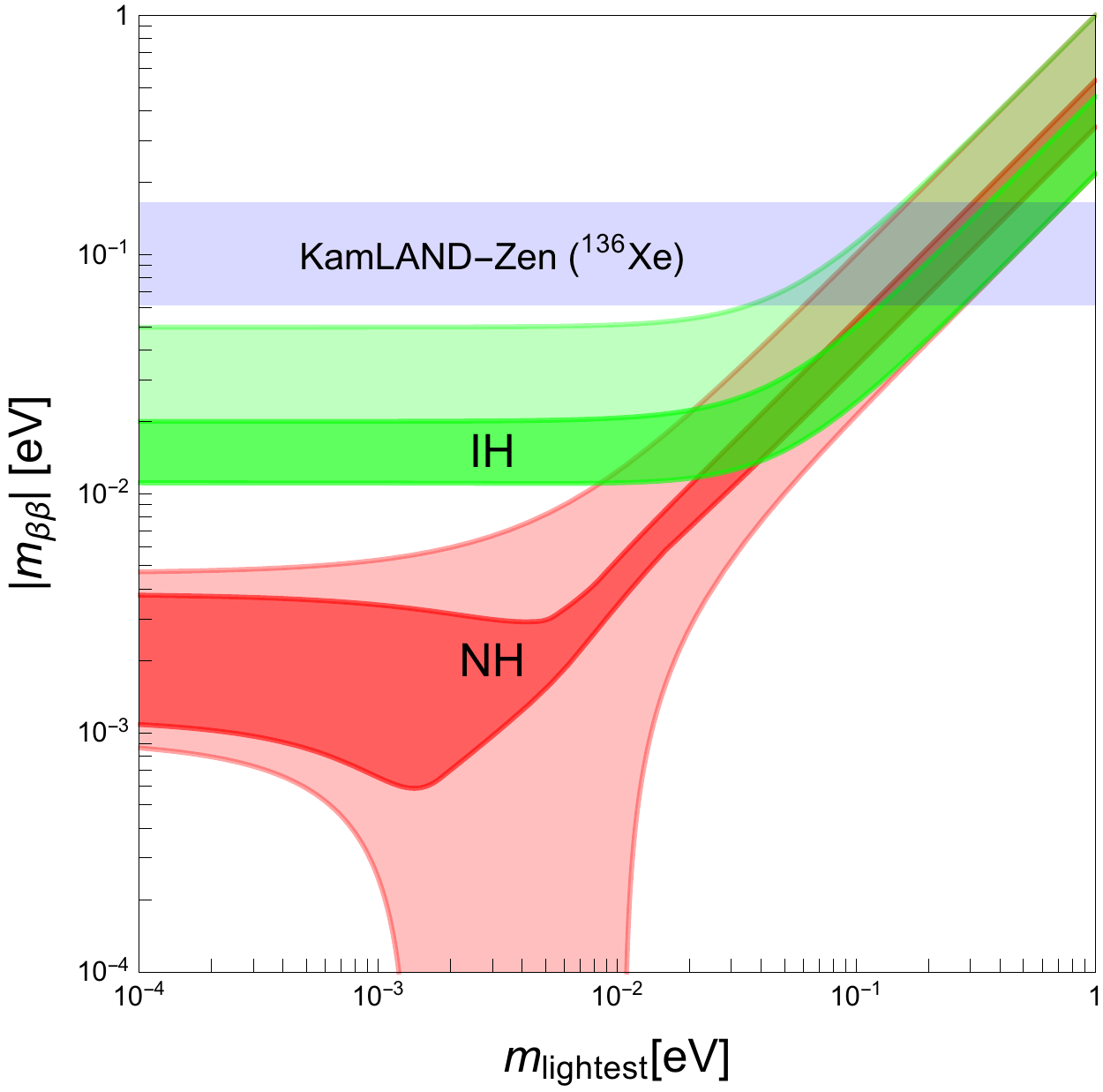}
\caption{Effective Majorana neutrino mass parameter $|m_{\protect\beta%
\protect\beta}|$ as a function of the lightest active neutrino mass $m_{%
\text{lightest}}$. The light shaded regions indicate the 3$\protect\sigma$
ranges calculated from the oscillation parameter uncertainties \protect\cite%
{Forero:2014bxa}, and the dark shaded areas correspond to the allowed values
predicted by the model. The horizontal blue band indicates 90\% C.L. upper
limits on $|m_{\protect\beta\protect\beta}|$ with ${}^{136}$Xe from
the KamLAND-Zen experiment~\protect\cite{KamLAND-Zen:2016pfg}. }
\label{mee}
\end{figure}

We show in Figure \ref{mee} the effective Majorana neutrino mass
parameter $|m_{\beta\beta}|$ versus the lightest active neutrino mass for
the cases of normal and inverted neutrino mass hierarchies. 
In order to determine the predicted ranges for $|m_{\beta\beta}|$ in
our model, we have randomly generated the angles $\omega_{23}$,
$\omega_{13}$ and $\psi$, as well as the light active neutrino mass
scale $m_{\nu}=\frac{\lambda_{2}v_{\xi }}{\sqrt{2+r^{2}}}$ in a range
of values where the neutrino mass squared splittings and the leptonic
mixing parameters are consistent with the observed neutrino
oscillation data.
Our predicted range of values for the effective Majorana neutrino mass
parameter has a lower bound, even in the case of normal hierarchy,
indicating that a complete destructive interference among the three
light neutrinos is prevented by our symmetry and the current
oscillation data.

The corresponding \nbb decay rates are within the reach of the
next-generation bolometric CUORE experiment \cite{Alessandria:2011rc}
or, more realistically, of the next-to-next-generation ton-scale
$0\nu \beta \beta $-decay experiments. It is worth mentioning that the
Majorana neutrino mass parameter has an upper bound on
$|m_{\beta\beta}|\leq (61-165)$ meV at 90\% C.L, as indicated by the
KamLAND-Zen experiment from the limit on the ${}^{136}\mathrm{Xe}$ $%
0\nu\beta\beta$
decay half-life $T^{0\nu\beta\beta}_{1/2}({}^{136}\mathrm{Xe}%
)\geq 1.07\times 10^{26}$
yr \cite{KamLAND-Zen:2016pfg}. This bound will be improved within a
not too far future. The GERDA \textquotedblleft
phase-II\textquotedblright experiment
\cite{Abt:2004yk,Ackermann:2012xja} is expected to reach
\mbox{$T^{0\nu\beta\beta}_{1/2}(^{76}{\rm Ge})\geq 2\times 10^{26}$
  yr}, corresponding to $|m_{\beta\beta}|\leq 100$ meV. A bolometric
CUORE experiment, using ${}^{130}\mathrm{Te}$
\cite{Alessandria:2011rc}, is currently under construction and its
estimated sensitivity is close to about
$T_{1/2}^{0\nu \beta \beta }(^{130}\mathrm{Te})\sim 10^{26}$ yr,
implying \mbox{$|m_{\beta\beta}|\leq 50$ meV.} Furthermore, 
there are plans for ton-scale next-to-next generation
$0\nu \beta \beta $ experiments with $^{136}$Xe
\cite{KamLANDZen:2012aa,Albert:2014fya} and $^{76}$Ge
\cite{Abt:2004yk,Guiseppe:2011me}, asserting sensitivities over
$T_{1/2}^{0\nu \beta \beta }\sim 10^{27}$ yr, corresponding to
$|m_{\beta\beta}|\sim 12-30$ meV.

\section{Discussions and Conclusions}
\label{sec:Discussions} %\label{sec:conclusions}

We have proposed a realistic extension of the standard model within
the Pati-Salam framework. The theory incorporates a flavor symmetry
based on the $\Delta \left( 27\right)$ group and realizes a realistic
Froggatt-Nielsen picture of quark mixing. 
Concerning the lepton sector, neutrino masses arise from an inverse
seesaw mechanism and the allowed ranges for the atmospheric mixing
angle $\theta_{23}$ and CP violating phase $\delta_{CP}$ are rather
restricted once one takes into account the precise measurements of the
remaining oscillation parameters. This makes the model rather
predictive.
Our main neutrino oscillation results are summarized in
Figs.~\ref{NHplot} and \ref{IHplot}. We find that, for normal neutrino
mass ordering, the atmospheric angle must lie in the higher octant and
that CP must be violated in oscillations. In contrast, for inverse
hierarchy, the lower octant is favored and the range of allowed
Jarlskog invariant extends from zero up to a non-maximal value.  Our
results concerning \nbb decay are summarized in Fig.\ref{mee}. They
indicate the existence of a lower bound for the \nbb decay rate, a
feature also encountered in other flavor
models~\cite{Dorame:2011eb,Dorame:2012zv,King:2013hj,Bonilla:2014xla,Gehrlein:2016wlc}.
As mentioned, neutrino masses arise from a low-scale seesaw mechanism,
whose messengers may be produced at the LHC either through a charged
or neutral gauge
portal~\cite{Das:2012ii,Deppisch:2013cya,Queiroz:2016gif}.
Admittedly, the model is rather complex, especially in its scalar
sector. However, it serves as a ``proof-of-concept'' attempt, namely,
to our knowledge this is the first time that a fully realistic and to
some extent predictive flavor realization of a Pati-Salam scenario is
given.

\section*{Acknowledgments}

This research is supported by the Spanish grants FPA2014-58183-P, Multidark
CSD2009-00064, SEV-2014-0398 (MINECO) and PROMETEOII/2014/084 (Generalitat
Valenciana); Chilean grants Fondecyt No. 1170803, No. 1150792 and CONICYT
PIA/Basal FB0821 and ACT1406; and Mexican grant CONACYT No. 274397. A.E.C.H is very grateful to Professor Hoang
Ngoc Long and the other members of its group for the warm hospitality of the
Institute of Physics, Vietnam Academy of Science and Technology, where this
work was completed.

\section*{Appendix}

\appendix

\section{Product rules of the $\Delta (27)$ discrete group}

\label{sec:product-rules-delta}

The $\Delta (27)$ discrete group is a subgroup of $SU(3)$, has 27 elements
divided into 11 conjugacy classes, so that it has 11 irreducible
representations. These irreducible representations are: two triplets, i.e., $%
\mathbf{3}_{[0][1]}$ (which we denote by $\mathbf{3}$)\ and its conjugate $%
\mathbf{3}_{[0][2]}$ (which we denote by $\overline{\mathbf{3}}$) and 9
singlets, i.e., $\mathbf{1}_{k,l}$ ($k,l=0,1,2$), where $k$ and $l$
correspond to the $Z_{3}$ and $Z_{3}^{\prime }$ charges, respectively \cite%
{Ishimori:2010au}. The $\Delta (27)$ discrete group is a simple group of the
type $\Delta (3n^{2})$ with $n=3$ and is isomorphic to the semi-direct
product group $(Z_{3}^{\prime }\times Z_{3}^{\prime \prime })\rtimes Z_{3}$ 
\cite{Ishimori:2010au}. Indeed,
the simplest group of the type $\Delta (3n^{2})$ is $\Delta (3)\equiv Z_{3}$. 
The next group is $\Delta (12)$, which is isomorphic to $A_{4}$. Thus, the 
$\Delta (27)$ discrete group is the next simplest nontrivial group of the type 
$\Delta (3n^{2})$. It is worth mentioning that one can write any element of
the $\Delta (27)$ discrete group as $b^{k}a^{m}{a^{\prime }}^{n}$, where $b$, 
$a$ and $a^{\prime }$ correspond to the generators of the $Z_{3}$, $%
Z_{3}^{\prime }$ and $Z_{3}^{\prime \prime }$ cyclic groups, respectively.
These generators fulfill the relations: 
\begin{eqnarray}
&&a^{3}=a^{\prime 3}=b^{3}=1,\hspace*{0.5cm}aa^{\prime }=a^{\prime }a, 
\notag \\
&&bab^{-1}=a^{-1}a^{\prime -1},\,\,ba^{\prime }b^{-1}=a,  \label{aapbrela}
\end{eqnarray}
The characters of the $\Delta (27)$ discrete group are shown in Table 
\ref{tab:delta-27}. Here $n$ is the number of elements, $h$ is the order of each
element, and $\omega =e^{\frac{2\pi i}{3}}=-\frac{1}{2}+i\frac{\sqrt{3}}{2}$
is the cube root of unity, which satisfies the relations $1+\omega +\omega
^{2}=0$ and $\omega ^{3}=1$. %\newpage
The conjugacy classes of $\Delta (27)$ are given by: 
\begin{equation*}
\begin{array}{ccc}
C_{1}: & \{e\}, & h=1, \\ 
C_{1}^{(1)}: & \{a,a^{\prime 2}\}, & h=3, \\ 
C_{1}^{(2)}: & \{a^{2},a^{\prime }\}, & h=3, \\ 
C_{3}^{(0,1)}: & \{a^{\prime 2}a^{\prime 2}\}, & h=3, \\ 
C_{3}^{(0,2)}: & \{a^{\prime 2},a^{2},aa^{\prime }\}, & h=3, \\ 
C_{3}^{(1,p)}: & \{ba^{p},ba^{p-1}a^{\prime p-2}a^{\prime 2}\}, & h=3, \\ 
C_{3}^{(2,p)}: & \{ba^{p},ba^{p-1}a^{\prime p-2}a^{\prime 2}\}, & h=3. \\ 
&  & 
\end{array}%
\end{equation*}

\begin{table}[t]
\begin{center}
\begin{tabular}{|c|c|c|c|c|}
\hline
& h & $\chi_{1_{(r,s)}}$ & $\chi_{3_{[0,1]}}$ & $\chi_{3_{[0,2]}}$ \\ \hline
$1C_1$ & 1 & 1 & 3 & 3 \\ \hline
$1C_1^{(1)}$ & 1 & 1 & $3\omega^2$ & $3\omega$ \\ \hline
$1C_1^{(2)}$ & 1 & 1 & $3\omega$ & $3\omega^2$ \\ \hline
$3C_1^{(0,1)}$ & $3$ & $\omega^{s}$ & $0$ & $0$ \\ \hline
$3C_1^{(0,2)}$ & $3$ & $\omega^{2s}$ & $0$ & $0$ \\ \hline
$C_3^{(1,p)}$ & $3$ & $\omega^{r+s p}$ & 0 & 0 \\ \hline
$C_3^{(2,p)}$ & $3$ & $\omega^{2r+s p}$ & 0 & 0 \\ \hline
\end{tabular}%
\end{center}
\caption{Characters of $\Delta (27)$}
\label{tab:delta-27}
\end{table}
The tensor products between $\Delta (27)$ triplets are described by the
following relations \cite{Ishimori:2010au}: 
\begin{eqnarray}
\vspace{-1cm}%
\begin{pmatrix}
x_{1,-1} \\ 
x_{0,1} \\ 
x_{-1,0} \\ 
\end{pmatrix}
_{\mathbf{3}}\otimes 
\begin{pmatrix}
y_{1,-1} \\ 
y_{0,1} \\ 
y_{-1,0} \\ 
\end{pmatrix}%
_{\mathbf{3}} &=&%
\begin{pmatrix}
x_{1,-1}y_{1,-1} \\ 
x_{0,1}y_{0,1} \\ 
x_{-1,0}y_{-1,0} \\ 
\end{pmatrix}%
_{\mathbf{\bar{3}}_{S_{1}}}\oplus \frac{1}{2}
\begin{pmatrix}
x_{0,1}y_{-1,0}+x_{-1,0}y_{0,1} \\ 
x_{-1,0}y_{1,-1}+x_{1,-1}y_{-1,0} \\ 
x_{1,-1}y_{0,1}+x_{0,1}y_{1,-1} \\ 
\end{pmatrix}
_{\mathbf{\bar{3}}_{S_{2}}}  \notag \\
&&\oplus \frac{1}{2}%
\begin{pmatrix}
x_{0,1}y_{-1,0}-x_{-1,0}y_{0,1} \\ 
x_{-1,0}y_{1,-1}-x_{1,-1}y_{-1,0} \\ 
x_{1,-1}y_{0,1}-x_{0,1}y_{1,-1} \\ 
\end{pmatrix}
_{\mathbf{\bar{3}}_{A}}, \\
\begin{pmatrix}
x_{2,-2} \\ 
x_{0,2} \\ 
x_{-2,0} \\ 
\end{pmatrix}
_{\mathbf{\bar{3}}}\otimes 
\begin{pmatrix}
y_{2,-2} \\ 
y_{0,2} \\ 
y_{-2,0} \\ 
\end{pmatrix}%
_{\mathbf{\bar{3}}}
&=&%
\begin{pmatrix}
x_{2,-2}y_{2,-2} \\ 
x_{0,2}y_{0,2} \\ 
x_{-2,0}y_{-2,0} \\ 
\end{pmatrix}%
_{\mathbf{3}_{S_{1}}}\oplus \frac{1}{2}
\begin{pmatrix}
x_{0,2}y_{-2,0}+x_{-2,0}y_{0,2} \\ 
x_{-2,0}y_{2,-2}+x_{2,-2}y_{-2,0} \\ 
x_{2,-2}y_{0,2}+x_{0,2}y_{2,-2} \\ 
\end{pmatrix}%
_{\mathbf{3}_{S_{2}}}  \notag \\
&&\oplus \frac{1}{2}%
\begin{pmatrix}
x_{0,2}y_{-2,0}-x_{-2,0}y_{0,2} \\ 
x_{-2,0}y_{2,-2}-x_{2,-2}y_{-2,0} \\ 
x_{2,-2}y_{0,2}-x_{0,2}y_{2,-2} \\ 
\end{pmatrix}%
_{\mathbf{3}_{A}}, \\
\begin{pmatrix}
x_{1,-1} \\ 
x_{0,1} \\ 
x_{-1,0} \\ 
\end{pmatrix}%
_{\mathbf{3}}\otimes 
\begin{pmatrix}
y_{-1,1} \\ 
y_{0,-1} \\ 
y_{1,0} \\ 
\end{pmatrix}%
_{\mathbf{\bar{3}}} 
&=&\sum_{r}(x_{1,-1}y_{-1,1}+\omega
^{2r}x_{0,1}y_{0,-1}+\omega ^{r}x_{-1,0}y_{1,0})_{\mathbf{1}_{(r,0)}}  \notag
\\
&\oplus &\sum_{r}(x_{1,-1}y_{0,-1}+\omega ^{2r}x_{0,1}y_{1,0}+\omega
^{r}x_{-1,0}y_{-1,1})_{\mathbf{1}_{(r,1)}}  \notag \\
&\oplus &\sum_{r}(x_{1,-1}y_{1,0}+\omega ^{2r}x_{0,1}y_{-1,1}+\omega
^{r}x_{-1,0}y_{0,-1})_{\mathbf{1}_{(r,2)}},  \notag \\
&&
\end{eqnarray}%
where we introduced the shorthand notation $\mathbf{3}_{[0][1]}\equiv \mathbf{3}$ and $\mathbf{3}_{[0][2]}\equiv \mathbf{\bar{3}}$ used in Eq.(\ref{eq:Yukawa-1}). In the above formulas $\mathbf{3}_{A}$ and $\mathbf{3}_{S_{1,2}}$ are an antisymmetric and two variants of symmetric triplets. The multiplication rules between $\Delta (27)$ singlets and $\Delta (27)$
triplets are given by \cite{Ishimori:2010au}: 
\begin{eqnarray}
&&%
\begin{pmatrix}
x_{(1,-1)} \\ 
x_{(0,1)} \\ 
x_{(-1,0)}%
\end{pmatrix}%
_{\mathbf{3}_{[0][1]}}\otimes (z)_{1_{k,l}}=%
\begin{pmatrix}
x_{(1,-1)}z \\ 
\omega ^{r}x_{(0,1)}z \\ 
\omega ^{2r}x_{(-1,0)}z%
\end{pmatrix}%
_{\mathbf{3}_{[l][1+l]}}, \\
&&%
\begin{pmatrix}
x_{(2,-2)} \\ 
x_{(0,2)} \\ 
x_{(-2,0)}%
\end{pmatrix}%
_{\mathbf{3}_{[0][2]}}\otimes (z)_{1_{k,l}}=%
\begin{pmatrix}
x_{(2,-2)}z \\ 
\omega ^{r}x_{(0,2)}z \\ 
\omega ^{2r}x_{(-2,0)}%
\end{pmatrix}%
_{\mathbf{3}_{[l][2+l]}}.
\end{eqnarray}%
The tensor products of $\Delta (27)$ singlets $\mathbf{1}_{k,\ell }$ and $%
\mathbf{1}_{k^{\prime },\ell ^{\prime }}$ take the form \cite%
{Ishimori:2010au}: 
\begin{equation}
\mathbf{1}_{k,\ell }\otimes \mathbf{1}_{k^{\prime },\ell ^{\prime }}=\mathbf{%
1}_{k+k^{\prime }\mathrm{\, mod \, }3,\ell +\ell ^{\prime }\mathrm{ \,mod\, }3}.
\end{equation}%
From the equation given above, we obtain explicitly the singlet
multiplication rules of the $\Delta (27)$ group, which are given in Table %
\ref{D27multiplets}.

\begin{table*}[tbh]
\begin{center}
\begin{tabular}{|c||c|c|c|c|c|c|c|c|}
\hline
Singlets & ~ $\mathbf{1}_{01}$ ~ & ~ $\mathbf{1}_{02}$ ~ & ~ $\mathbf{1}%
_{10} $ ~ & ~ $\mathbf{1}_{11}$~ & ~ $\mathbf{1}_{12}$ ~ & ~ $\mathbf{1}%
_{20} $ ~ & ~ $\mathbf{1}_{21}$~ & $\mathbf{1}_{22}$ \\ \hline\hline
$\mathbf{1}_{01}$ & $\mathbf{1}_{02}$ & $\mathbf{1}_{00}$ & $\mathbf{1}_{11}$
& $\mathbf{1}_{12}$ & $\mathbf{1}_{10}$ & $\mathbf{1}_{21}$ & $\mathbf{1}%
_{22}$ & $\mathbf{1}_{20}$ \\ \hline
$\mathbf{1}_{02}$ & $\mathbf{1}_{00}$ & $\mathbf{1}_{01}$ & $\mathbf{1}_{12}$
& $\mathbf{1}_{10}$ & $\mathbf{1}_{11}$ & $\mathbf{1}_{22}$ & $\mathbf{1}%
_{20}$ & $\mathbf{1}_{21}$ \\ \hline
$\mathbf{1}_{10}$ & $\mathbf{1}_{11}$ & $\mathbf{1}_{12}$ & $\mathbf{1}_{20}$
& $\mathbf{1}_{21}$ & $\mathbf{1}_{22}$ & $\mathbf{1}_{00}$ & $\mathbf{1}%
_{01}$ & $\mathbf{1}_{02}$ \\ \hline
$\mathbf{1}_{11}$ & $\mathbf{1}_{12}$ & $\mathbf{1}_{10}$ & $\mathbf{1}_{21}$
& $\mathbf{1}_{22}$ & $\mathbf{1}_{20}$ & $\mathbf{1}_{01}$ & $\mathbf{1}%
_{02}$ & $\mathbf{1}_{00}$ \\ \hline
$\mathbf{1}_{12}$ & $\mathbf{1}_{10}$ & $\mathbf{1}_{11}$ & $\mathbf{1}_{22}$
& $\mathbf{1}_{20}$ & $\mathbf{1}_{21}$ & $\mathbf{1}_{02}$ & $\mathbf{1}%
_{00}$ & $\mathbf{1}_{01}$ \\ \hline
$\mathbf{1}_{20}$ & $\mathbf{1}_{21}$ & $\mathbf{1}_{22}$ & $\mathbf{1}_{00}$
& $\mathbf{1}_{01}$ & $\mathbf{1}_{02}$ & $\mathbf{1}_{10}$ & $\mathbf{1}%
_{11}$ & $\mathbf{1}_{12}$ \\ \hline
$\mathbf{1}_{21}$ & $\mathbf{1}_{22}$ & $\mathbf{1}_{20}$ & $\mathbf{1}_{01}$
& $\mathbf{1}_{02}$ & $\mathbf{1}_{00}$ & $\mathbf{1}_{11}$ & $\mathbf{1}%
_{12}$ & $\mathbf{1}_{10}$ \\ \hline
$\mathbf{1}_{22}$ & $\mathbf{1}_{20}$ & $\mathbf{1}_{21}$ & $\mathbf{1}_{02}$
& $\mathbf{1}_{00}$ & $\mathbf{1}_{01}$ & $\mathbf{1}_{12}$ & $\mathbf{1}%
_{10}$ & $\mathbf{1}_{11}$ \\ \hline
\end{tabular}%
\end{center}
\caption{The singlet multiplications of the group $\Delta (27)$.}
\label{D27multiplets}
\end{table*}

\section{Scalar potential for one $\Delta (27)$ scalar triplet}

\label{sec:scalarpotential1triplet} The scalar potential for a $\Delta (27)$
scalar triplet, i.e., $\xi $ is given by: 
\begin{eqnarray}
V &=&-\mu_{\xi }^{2}\left( \xi \xi ^{\ast }\right)_{\mathbf{\mathbf{1}_{0%
\mathbf{,0}}}}+\kappa_{1}\left( \xi \xi ^{\ast }\right)_{\mathbf{\mathbf{1}%
_{0\mathbf{,0}}}}\left( \xi \xi ^{\ast }\right)_{\mathbf{\mathbf{1}_{0%
\mathbf{,0}}}}+\kappa_{2}\left( \xi \xi ^{\ast }\right)_{\mathbf{\mathbf{1}%
_{1\mathbf{,0}}}}\left( \xi \xi ^{\ast }\right)_{\mathbf{\mathbf{1}_{2%
\mathbf{,0}}}}+\kappa_{3}\left( \xi \xi ^{\ast }\right)_{\mathbf{\mathbf{1}%
_{0,1}}}\left( \xi \xi ^{\ast }\right)_{\mathbf{\mathbf{1}_{0,2}}}  \notag
\\
&&+\kappa_{4}\left[ \left( \xi \xi ^{\ast }\right)_{\mathbf{\mathbf{1}%
_{1,1}}}\left( \xi \xi ^{\ast }\right)_{\mathbf{\mathbf{1}_{2,2}}}+\text{h.c.}%
\right] +\kappa_{5}\left( \xi \xi \right)_{\overline{\mathbf{3}}\mathbf{%
_{S_{1}}}}\left( \xi ^{\ast }\xi ^{\ast }\right)_{\mathbf{3}%
_{S_{1}}}+\kappa _{6}\left( \xi \xi \right) _{\overline{\mathbf{3}}\mathbf{%
_{S_{2}}}}\left( \xi ^{\ast }\xi ^{\ast }\right) _{\mathbf{3}_{S_{2}}} 
\notag \\
&&+\kappa _{7}\left[ \left( \xi \xi \right) _{\overline{\mathbf{3}}\mathbf{%
_{S_{1}}}}\left( \xi ^{\ast }\xi ^{\ast }\right) _{\mathbf{3}_{S_{2}}}+\text{h.c.}
\right]  \label{potential1triplet}
\end{eqnarray}

Assuming the VEV configuration for the $\Delta (27)$ scalar triplet $\xi $:
\begin{equation}
\left\langle \xi \right\rangle =\frac{v_{\xi }}{\sqrt{2+r^{2}}}\left(
r,e^{-i\psi },e^{i\psi }\right) ,  \label{VEV1triplet}
\end{equation}

We find that the scalar potential minimization equations are: 
\begin{eqnarray}
\frac{\partial V}{\partial \xi\text{$_{_{1}}$}} &=&\frac{2v_{\xi }^{3}\left\{
\left( 2\kappa _{3}+\kappa _{4}+2\kappa _{7}\right) \cos \psi +r\left[
2\kappa _{3}+\kappa _{4}+2\kappa _{7}+2\left( \kappa _{1}+\kappa _{2}+\kappa
_{5}\right) r^{2}\right] \right\} }{\left( r^{2}+2\right) ^{3/2}}  \notag \\
&&+\frac{2v_{\xi }\left[ 2\left( 2\kappa _{1}-\kappa _{2}+\kappa _{3}-\kappa
_{4}+\kappa _{6}\right) rv_{\xi }^{2}\cos 2\psi -\mu _{\xi }^{2}r\left(
r^{2}+2\right) \right] }{\left( r^{2}+2\right) ^{3/2}}  \notag \\
&=&0\,  \notag \\
\frac{\partial V}{\partial \xi\text{$_{2}$}} &=&\frac{e^{-3i\psi }v_{\xi }^{3}%
\left[ 4\left( \kappa _{1}+\kappa _{2}+\kappa _{5}\right) +2\left( 2\kappa
_{1}-\kappa _{2}+\kappa _{3}-\kappa _{4}+\kappa _{6}\right) r^{2}e^{2i\psi }%
\right] }{\left( r^{2}+2\right) ^{3/2}}  \notag \\
&&+\frac{e^{i\psi }v_{\xi }^{3}\left[ 4\kappa _{1}-2\kappa _{2}+2\kappa
_{6}+2\kappa _{7}r^{2}+2\kappa _{3}\left( r^{2}+1\right) +\kappa _{4}\left(
r^{2}-2\right) \right] }{\left( r^{2}+2\right) ^{3/2}}  \notag \\
&&+\frac{e^{-3i\psi }\left\{ v_{\xi }^{3}\left[ 2\left( 2\kappa _{3}+\kappa
_{4}+2\kappa _{7}\right) re^{3i\psi }+\left( 2\kappa _{3}+\kappa
_{4}+2\kappa _{7}\right) re^{5i\psi }\right] -2\mu _{\xi }^{2}\left(
r^{2}+2\right) e^{2i\psi }v_{\xi }\right\} }{\left( r^{2}+2\right) ^{3/2}} 
\notag \\
&=&0,\hspace*{0.5cm}\hspace*{0.5cm}\hspace*{0.5cm}\hspace*{0.5cm}  \notag \\
\frac{\partial V}{\partial \xi\text{$_{3}$}} &=&\frac{e^{-2i\psi }v_{\xi }^{3}%
\left[ 4\left( \kappa _{1}+\kappa _{2}+\kappa _{5}\right) e^{5i\psi
}+2\left( 2\kappa _{1}-\kappa _{2}+\kappa _{3}-\kappa _{4}+\kappa
_{6}\right) r^{2}e^{3i\psi }\right] }{\left( r^{2}+2\right) ^{3/2}}  \notag
\\
&&+\frac{e^{-2i\psi }v_{\xi }^{3}\left[ e^{i\psi }\left( 4\kappa
_{1}-2\kappa _{2}+2\kappa _{6}+2\kappa _{7}r^{2}+2\kappa _{3}\left(
r^{2}+1\right) +\kappa _{4}\left( r^{2}-2\right) \right) \right] }{\left(
r^{2}+2\right) ^{3/2}}  \notag \\
&&+\frac{e^{-2i\psi }\left\{ v_{\xi }^{3}\left[ 2\left( 2\kappa _{3}+\kappa
_{4}+2\kappa _{7}\right) re^{2i\psi }+\left( 2\kappa _{3}+\kappa
_{4}+2\kappa _{7}\right) r\right] -2\mu _{\xi }^{2}\left( r^{2}+2\right)
e^{3i\psi }v_{\xi }\right\} }{\left( r^{2}+2\right) ^{3/2}}  \notag \\
&=&0.
\end{eqnarray}

Then, from the scalar potential minimization equations, we find the
following relations: 
\begin{eqnarray}
\mu _{\xi }^{2} &=&\frac{v_{\xi }^{2}}{r\left( r^{2}+2\right) }\bigg\{ \left( 2\kappa _{3}+\kappa _{4}+2\kappa
_{7}\right) \cos \psi +r\left[ 2\kappa _{3}+\kappa _{4}+2\kappa _{7}+2\left(
\kappa _{1}+\kappa _{2}+\kappa _{5}\right) r^{2}\right]\notag\\&& +2\left( 2\kappa
_{1}-\kappa _{2}+\kappa _{3}-\kappa _{4}+\kappa _{6}\right) r\cos 2\psi %
\bigg\} ,  \notag \\
&&2\left[ -6\kappa _{2}-4\kappa _{5}+2\kappa _{6}+2\kappa _{7}r^{2}+2\kappa
_{3}\left( r^{2}+1\right) +\kappa _{4}\left( r^{2}-2\right) \right] \sin
\psi \cos \psi  \notag \\
&&+\left[ 2\left( 2\kappa _{3}+\kappa _{4}+2\kappa _{7}\right) r\cos 2\psi
+3\left( 2\kappa _{3}+\kappa _{4}+2\kappa _{7}\right) r\right] \sin \psi =0,
\notag \\
&&\left( 2\kappa _{3}+\kappa _{4}+2\kappa _{7}\right) \left( r^{2}-1\right)
\cos 2\psi \notag\\&&-r\left\{ 2\left( 2\kappa _{3}+\kappa _{4}+2\kappa _{7}\right) + 
\left[ 6\kappa _{2}-4\kappa _{3}+\kappa _{4}+4\kappa _{5}-2\left( \kappa
_{6}+\kappa _{7}\right) \right] r^{2}\right\} \cos \psi  \notag \\
&&+\left( 2\kappa _{3}+\kappa _{4}+2\kappa _{7}\right) \left(
2r^{2}-1\right) +2\left( 3\kappa _{2}-\kappa _{3}+\kappa _{4}+2\kappa
_{5}-\kappa _{6}\right) r\cos 3\psi =0.
\end{eqnarray} 
Thus, the VEV pattern for the $\Delta (27)$ triplet, i.e., $\xi$ given by
Eq. (\ref{VEV1triplet}), is compatible with a global minimum of the scalar
potential of Eq. (\ref{potential1triplet}) for a large region of parameter
space.

\section{Scalar potential for three $\Delta (27)$ scalar triplets}

\label{sec:scalarpotential3triplets} 

The scalar potential for the three $\Delta
(27)$ scalar triplets  $\rho$, $\tau$ and $\eta$  is 
\begin{equation}
V=V_{\rho}+V_{\tau}+V_{\eta}+V_{\rho,\tau}+V_{\tau,\eta}+V_{\rho,\eta}\,,  \label{scalarpotential3triplets}
\end{equation}%
where $V_{\rho}$, $V_{\tau}$ and $V_{\eta}$ are the scalar potentials for the $\Delta
(27)$ scalar triplets $\rho$, $\tau$ and $\eta$, respectively, whereas $V_{\rho,\tau}$, $%
V_{\tau,\eta}$ and $V_{\rho,\eta}$ describe interaction terms involving the pairs ($\rho$, $\tau$), ($\tau$, $\eta$) and ($\rho$, $\eta$). The
different parts of the scalar potential for the three $\Delta (27)$ scalar
triplets take the form: 
\begin{eqnarray}
V_{\rho} &=&-\mu _{\rho}^{2}\left( \rho\rho^{\ast }\right) _{\mathbf{\mathbf{1}_{0%
\mathbf{,0}}}}+\kappa _{\rho,1}\left( \rho\rho^{\ast }\right) _{\mathbf{\mathbf{1}_{0%
\mathbf{,0}}}}\left( \rho\rho^{\ast }\right) _{\mathbf{\mathbf{1}_{0\mathbf{,0}}}%
}+\kappa _{\rho,2}\left( \rho\rho^{\ast }\right) _{\mathbf{\mathbf{1}_{1\mathbf{,0}}}%
}\left( \rho\rho^{\ast }\right) _{\mathbf{\mathbf{1}_{2\mathbf{,0}}}}  \notag \\
&&+\kappa
_{\rho,3}\left( \rho\rho^{\ast }\right) _{\mathbf{\mathbf{1}_{0,1}}}\left( \rho\rho^{\ast
}\right) _{\mathbf{\mathbf{1}_{0,2}}}+\kappa _{\rho,4}\left[ \left( \rho\rho^{\ast }\right) _{\mathbf{\mathbf{1}_{1,1}}%
}\left( \rho\rho^{\ast }\right) _{\mathbf{\mathbf{1}_{2,2}}}+\text{h.c.}\right] +\kappa
_{\rho,5}\left( \rho\rho\right) _{\overline{\mathbf{3}}\mathbf{_{S_{1}}}}\left(
\rho^{\ast }\rho^{\ast }\right) _{\mathbf{3}_{S_{1}}} \notag \\
&&+\kappa _{\rho,6}\left(
\rho\rho\right) _{\overline{\mathbf{3}}\mathbf{_{S_{2}}}}\left( \rho^{\ast }\rho^{\ast
}\right) _{\mathbf{3}_{S_{2}}} +\kappa _{\rho,7}\left[ \left( \rho\rho\right) _{\overline{\mathbf{3}}\mathbf{%
_{S_{1}}}}\left( \rho^{\ast }\rho^{\ast }\right) _{\mathbf{3}_{S_{2}}}+\text{h.c.}\right]\,,\\
V_{\tau}&=&V_{\rho}\left( \rho \to \tau,\mu _{\rho} \to \mu _{\tau},\kappa _{\rho,j} \to \kappa
_{\tau,j}\right) \,,\\
V_{\eta}&=&V_{\rho}\left( \rho \rightarrow \eta,\mu _{\rho} \to \mu _{\eta},\kappa _{\rho,j} \to
\kappa _{\eta,j}\right) \,,\\
V_{\rho,\tau} &=&\gamma _{\rho\tau,1}\left( \rho\rho^{\ast }\right) _{\mathbf{\mathbf{1}_{0%
\mathbf{,0}}}}\left( \tau\tau^{\ast }\right) _{\mathbf{\mathbf{1}_{0\mathbf{,0}}}%
}+\kappa _{\rho\tau,1}\left( \rho\tau^{\ast }\right) _{\mathbf{\mathbf{1}_{0\mathbf{,0}}}%
}\left( \rho^{\ast }\tau\right) _{\mathbf{\mathbf{1}_{0\mathbf{,0}}}}+\gamma
_{\rho\tau,2}\left[ \left( \rho\rho^{\ast }\right) _{\mathbf{\mathbf{1}_{1\mathbf{,0}}}%
}\left( \tau\tau^{\ast }\right) _{\mathbf{\mathbf{1}_{2\mathbf{,0}}}}+\text{h.c.}\right] 
\notag \\
&&+\kappa _{\rho\tau,2}\left[ \left( \rho\tau^{\ast }\right) _{\mathbf{\mathbf{1}_{1%
\mathbf{,0}}}}\left( \rho\tau^{\ast }\right) _{\mathbf{\mathbf{1}_{2\mathbf{,0}}}%
}+\text{h.c.}\right] +\gamma _{\rho\tau,3}\left[ \left( \rho\rho^{\ast }\right) _{\mathbf{%
\mathbf{1}_{0,1}}}\left( \tau\tau^{\ast }\right) _{\mathbf{\mathbf{1}_{0,2}}}+\text{h.c.}%
\right]  \notag \\
&&+\kappa _{\rho\tau,3}\left[ \left( \rho\tau^{\ast }\right) _{\mathbf{\mathbf{1}_{0,1}}%
}\left( \rho\tau^{\ast }\right) _{\mathbf{\mathbf{1}_{0,2}}}+\text{h.c.}\right] +\gamma
_{\rho\tau,4}\left[ \left( \rho\rho^{\ast }\right) _{\mathbf{\mathbf{1}_{1,1}}}\left(
\tau\tau^{\ast }\right) _{\mathbf{\mathbf{1}_{2,2}}}+\text{h.c.}\right]  \notag \\
&&+\kappa _{\rho\tau,4}\left[ \left( \rho\tau^{\ast }\right) _{\mathbf{\mathbf{1}_{1,1}}%
}\left( \rho\tau^{\ast }\right) _{\mathbf{\mathbf{1}_{2,2}}}+\text{h.c.}\right] +\gamma
_{\rho\tau,5}\left[ \left( \rho\rho\right) _{\overline{\mathbf{3}}\mathbf{_{S_{1}}}%
}\left( \tau^{\ast }\tau^{\ast }\right) _{\mathbf{3}_{S_{1}}}+\text{h.c.}\right]  \notag \\
&&+\gamma _{\rho\tau,6}\left[ \left( \rho\rho\right) _{\overline{\mathbf{3}}\mathbf{%
_{S_{2}}}}\left( \tau^{\ast }\tau^{\ast }\right) _{\mathbf{3}_{S_{2}}}+\text{h.c.}\right]
+\kappa _{\rho\tau,5}\left( \rho\tau\right) _{\overline{\mathbf{3}}\mathbf{_{S_{1}}}%
}\left( \rho^{\ast }\tau^{\ast }\right) _{\mathbf{3}_{S_{1}}}+\kappa _{\rho\tau,6}\left(
\rho\tau\right) _{\overline{\mathbf{3}}\mathbf{_{S_{2}}}}\left( \rho^{\ast }\tau^{\ast
}\right) _{\mathbf{3}_{S_{2}}}  \notag \\
&&+\gamma _{\rho\tau,7}\left[ \left( \rho\rho\right) _{\overline{\mathbf{3}}\mathbf{%
_{S_{1}}}}\left( \tau^{\ast }\tau^{\ast }\right) _{\mathbf{3}_{S_{2}}}+\text{h.c.}\right]
+\kappa _{\rho\tau,7}\left[ \left( \rho\tau\right) _{\overline{\mathbf{3}}\mathbf{%
_{S_{1}}}}\left( \rho^{\ast }\tau^{\ast }\right) _{\mathbf{3}_{S_{2}}}+\text{h.c.}\right] 
\notag \\
&&+\kappa _{\rho\tau,8}\left[ \left( \rho\tau\right) _{\overline{\mathbf{3}}\mathbf{_{A}}%
}\left( \rho^{\ast }\tau^{\ast }\right) _{\mathbf{3}_{A}}+\text{h.c.}\right] +\kappa
_{\rho\tau,9}\left[ \left( \rho\tau\right) _{\overline{\mathbf{3}}\mathbf{_{A}}}\left(
\rho^{\ast }\tau^{\ast }\right) _{\mathbf{3}_{S_{1}}}+\text{h.c.}\right]  \notag \\
&&+\kappa _{\rho\tau,10}\left[ \left( \rho\tau\right) _{\overline{\mathbf{3}}\mathbf{_{A}%
}}\left( \rho^{\ast }\tau^{\ast }\right) _{\mathbf{3}_{S_{2}}}+\text{h.c.}\right]
\\
V_{\rho,\eta}&=&V_{\rho,\tau}\left( \tau \to \eta,\mu _{\tau} \to \mu _{\eta},\kappa _{\tau,j} \to \kappa
_{\eta,j}\right) \,,\\
V_{\tau,\eta}&=&V_{\rho,\eta}\left( \rho \to \tau,\mu _{\rho} \to \mu _{\tau},\kappa _{\rho,j} \to \kappa
_{\tau,j}\right)\, .
\end{eqnarray}%
Now we determine the conditions under which the VEV pattern for
the $\Delta \left( 27\right) $ scalar triplets is a solution of the scalar
potential given in Eq. (\ref{scalarpotential3triplets}). In view of the very
large number of parameters of the scalar potential for the $\Delta \left(
27\right) $ scalar triplets and in order to simplify the analysis, we assume
universality in its trilinear and quartic couplings, i.e. 
\begin{eqnarray}
\kappa _{\rho,i} &=&\kappa _{\tau,i}=\kappa _{\eta,i}=\kappa _{i},\hspace*{0.5cm}%
\hspace*{0.5cm}\gamma _{\rho\tau,i}=\gamma _{\rho\eta,i}=\gamma _{\tau\eta,i}=\gamma _{i},%
\hspace*{0.5cm}\hspace*{0.5cm}i=1,2,\cdots 7.  \notag \\
\kappa _{\rho\tau,j} &=&\kappa _{\rho\eta,j}=\kappa _{\tau\eta,j}=\lambda _{j},\hspace*{0.5cm}%
\hspace*{0.5cm}j=1,2,\cdots 10
\end{eqnarray}

Considering the VEV alignment
\begin{equation}
\left\langle \rho\right\rangle =\left( v_{\rho},0,0\right)  ,\hspace*{0.5cm}%
\hspace*{0.5cm}\left\langle \eta\right\rangle =\left( 0,v_{\eta},0\right) ,
\hspace*{0.5cm}\hspace*{%
0.5cm}\left\langle \tau\right\rangle =\left( 0,0,v_{\tau}\right),
\label{VEVpatternUWT}
\end{equation}
We find the following scalar potential minimization equations: 
\begin{eqnarray}
\frac{\partial V}{\partial \rho\text{$_{_{1}}$}} &=&\frac{v_{\rho}}{2}\left[ -4\mu
_{\rho}^{2}+8\left( \kappa _{1}+\kappa _{2}+\kappa _{5}\right) v_{\rho}^{2}-2\lambda
_{10}\left( v_{\tau}^{2}-v_{\eta}^{2}\right) +\left( 4\gamma _{1}-4\gamma _{2}+\lambda
_{6}+2\lambda _{8}\right) \left( v_{\tau}^{2}+v_{\eta}^{2}\right) \right] =0,  \notag \\
\frac{\partial V}{\partial \rho\text{$_{2}$}} &=&v_{\rho}v_{\tau}^{2}\left( 2\lambda
_{3}-\lambda _{4}\right) =0,\hspace*{0.5cm}\hspace*{0.5cm}\hspace*{0.5cm}%
\hspace*{0.5cm}\hspace*{0.5cm}\hspace*{0.5cm}\frac{\partial V}{\partial \rho%
\text{$_{3}$}}=v_{\rho}v_{\eta}^{2}\left( 2\lambda _{3}-\lambda _{4}\right) =0,  \notag \\
\frac{\partial V}{\partial \tau\text{$_{1}$}} &=&v_{\tau}v_{\eta}^{2}\left( 2\lambda
_{3}-\lambda _{4}\right) =0,\hspace*{0.5cm}\hspace*{0.5cm}\hspace*{0.5cm}%
\hspace*{0.5cm}\frac{\partial V}{\partial \tau\text{$_{2}$}}=2v_{\rho}^{2}v_{\tau}\left(
\gamma _{7}+\lambda _{3}+\lambda _{4}\right) =0,  \notag \\
\frac{\partial V}{\partial \tau\text{$_{3}$}} &=&\frac{v_{\tau}}{2}\left[ -4\mu
_{\tau}^{2}+8\left( \kappa _{1}+\kappa _{2}+\kappa _{5}\right) v_{\tau}^{2}+\left(
4\gamma _{1}-4\gamma _{2}+\lambda _{6}+2\lambda _{8}-2\lambda _{10}\right)
\left( v_{\rho}^{2}+v_{\eta}^{2}\right) \right]  \notag \\
&=&0.  \notag \\
\frac{\partial V}{\partial \eta\text{$_{1}$}} &=&2v_{\tau}^{2}v_{\eta}\left( \gamma
_{7}+\lambda _{3}+\lambda _{4}\right) =0,\hspace*{0.5cm}\hspace*{0.5cm}%
\hspace*{0.5cm}\hspace*{0.5cm}\frac{\partial V}{\partial \eta\text{$_{3}$}}%
=2v_{\rho}^{2}v_{\eta}\left( \gamma _{7}+\lambda _{3}+\lambda _{4}\right) =0,  \notag \\
\frac{\partial V}{\partial \eta\text{$_{2}$}} &=&\frac{v_{\eta}}{2}\left[ -4\mu
_{\eta}^{2}+8\left( \kappa _{1}+\kappa _{2}+\kappa _{5}\right) v_{\eta}^{2}+2\lambda
_{10}\left( v_{\rho}^{2}-v_{\tau}^{2}\right) +\left( 4\gamma _{1}-4\gamma _{2}+\lambda
_{6}+2\lambda _{8}\right) \left( v_{\rho}^{2}+v_{\tau}^{2}\right) \right] \notag \\
&=&0.
\end{eqnarray}%
Then, from the scalar potential minimization equations, we find the
following relations: 
\begin{eqnarray}
\lambda _{4} &=&2\lambda _{3},\hspace*{0.5cm}\hspace*{0.5cm}\gamma
_{7}=-\left( \lambda _{3}+\lambda _{4}\right) ,  \notag \\
\mu _{\rho}^{2} &=&2\left( \kappa _{1}+\kappa _{2}+\kappa _{5}\right) v_{\rho}^{2}-%
\frac{1}{2}\lambda _{10}\left( v_{\tau}^{2}-v_{\eta}^{2}\right) +\frac{1}{4}\left( 4\gamma
_{1}-4\gamma _{2}+\lambda _{6}+2\lambda _{8}\right) \left(
v_{\tau}^{2}+v_{\eta}^{2}\right) ,  \notag \\
\mu _{\tau}^{2} &=&2\left( \kappa _{1}+\kappa _{2}+\kappa _{5}\right) v_{\tau}^{2}+%
\frac{1}{4}\left( 4\gamma _{1}-4\gamma _{2}+\lambda _{6}+2\lambda
_{8}-2\lambda _{10}\right) \left( v_{\rho}^{2}+v_{\eta}^{2}\right) ,  \notag \\
\mu _{\eta}^{2} &=&2\left( \kappa _{1}+\kappa _{2}+\kappa _{5}\right) v_{\eta}^{2}+%
\frac{1}{2}\lambda _{10}\left( v_{\rho}^{2}-v_{\tau}^{2}\right) +\frac{1}{4}\left( 4\gamma
_{1}-4\gamma _{2}+\lambda _{6}+2\lambda _{8}\right) \left(
v_{\rho}^{2}+v_{\tau}^{2}\right) .
\end{eqnarray}
Consequently, the VEV patterns for the three $\Delta (27)$ triplet scalars given by Eq. (\ref{VEVpatternUWT})
 are compatible with a
global minimum of the scalar potential of Eq. (\ref{scalarpotential3triplets}%
) for a large region of parameter space.
\bibliographystyle{JHEP}
\bibliography{corfu,merged_Valle,newrefs,d27,Refsall}

\providecommand{\href}[2]{#2}\begingroup\raggedright\begin{thebibliography}{10}

\bibitem{Kajita:2016cak}
T.~Kajita, \emph{{Nobel Lecture: Discovery of atmospheric neutrino
  oscillations}},
  \href{http://dx.doi.org/10.1103/RevModPhys.88.030501}{\emph{Rev. Mod. Phys.}
  {\bf 88} (2016) 030501}.

\bibitem{McDonald:2016ixn}
A.~B. McDonald, \emph{{Nobel Lecture: The Sudbury Neutrino Observatory:
  Observation of flavor change for solar neutrinos}},
  \href{http://dx.doi.org/10.1103/RevModPhys.88.030502}{\emph{Rev. Mod. Phys.}
  {\bf 88} (2016) 030502}.

\bibitem{Valle:2015pba}
J.~W. Valle and J.~C. Romao, \emph{{Neutrinos in high energy and astroparticle
  physics}}.
\newblock John Wiley \& Sons, 2015.

\bibitem{Ishimori:2010au}
H.~Ishimori et~al., \emph{{Non-Abelian Discrete Symmetries in Particle
  Physics}}, {\emph{Prog. Theor. Phys. Suppl.} {\bf 183} (2010) 1--163},
  [\href{https://arxiv.org/abs/1003.3552}{{\tt 1003.3552}}].

\bibitem{Morisi:2012fg}
S.~Morisi and J.~W.~F. Valle, \emph{{Neutrino masses and mixing: a flavour
  symmetry roadmap}}, {\emph{Fortsch.Phys.} {\bf 61} (2013) 466--492},
  [\href{https://arxiv.org/abs/1206.6678}{{\tt 1206.6678}}].

\bibitem{King:2013eh}
S.~F. King and C.~Luhn, \emph{{Neutrino Mass and Mixing with Discrete
  Symmetry}}, {\emph{Rept.Prog.Phys.} {\bf 76} (2013) 056201},
  [\href{https://arxiv.org/abs/1301.1340}{{\tt 1301.1340}}].

\bibitem{pati:1974yy}
J.~C. Pati and A.~Salam, \emph{Lepton number as the fourth color}, {\emph{Phys.
  Rev.} {\bf D10} (1974) 275--289}.

\bibitem{Mohapatra:1986bd}
R.~N. Mohapatra and J.~W.~F. Valle, \emph{Neutrino mass and baryon-number
  nonconservation in superstring models}, {\emph{Phys. Rev.} {\bf D34} (1986)
  1642}.

\bibitem{Akhmedov:1995ip}
E.~K. Akhmedov, M.~Lindner, E.~Schnapka and J.~Valle, \emph{{Left-right
  symmetry breaking in NJL approach}},
  \href{http://dx.doi.org/10.1016/0370-2693(95)01504-3}{\emph{Phys.Lett.} {\bf
  B368} (1996) 270--280}, [\href{https://arxiv.org/abs/hep-ph/9507275}{{\tt
  hep-ph/9507275}}].

\bibitem{Akhmedov:1995vm}
E.~K. Akhmedov, M.~Lindner, E.~Schnapka and J.~Valle, \emph{{Dynamical
  left-right symmetry breaking}},
  \href{http://dx.doi.org/10.1103/PhysRevD.53.2752}{\emph{Phys.Rev.} {\bf D53}
  (1996) 2752--2780}, [\href{https://arxiv.org/abs/hep-ph/9509255}{{\tt
  hep-ph/9509255}}].

\bibitem{Malinsky:2005bi}
M.~Malinsky, J.~Romao and J.~Valle, \emph{{Novel supersymmetric SO(10) seesaw
  mechanism}},
  \href{http://dx.doi.org/10.1103/PhysRevLett.95.161801}{\emph{Phys.Rev.Lett.}
  {\bf 95} (2005) 161801}, [\href{https://arxiv.org/abs/hep-ph/0506296}{{\tt
  hep-ph/0506296}}].

\bibitem{Perez:2013osa}
P.~Fileviez~Perez and M.~B. Wise, \emph{{Low Scale Quark-Lepton Unification}},
  \href{http://dx.doi.org/10.1103/PhysRevD.88.057703}{\emph{Phys. Rev.} {\bf
  D88} (2013) 057703}, [\href{https://arxiv.org/abs/1307.6213}{{\tt
  1307.6213}}].

\bibitem{Froggatt:1978nt}
C.~D. Froggatt and H.~B. Nielsen, \emph{{Hierarchy of Quark Masses, Cabibbo
  Angles and CP Violation}}, {\emph{Nucl. Phys.} {\bf B147} (1979) 277}.

\bibitem{Deppisch:2013cya}
F.~F. Deppisch, N.~Desai and J.~W.~F. Valle, \emph{{Is charged lepton flavour
  violation a high energy phenomenon?}}, {\emph{Phys.Rev.} {\bf D89} (2014)
  051302(R)}, [\href{https://arxiv.org/abs/1308.6789}{{\tt 1308.6789}}].

\bibitem{AguilarSaavedra:2012fu}
J.~Aguilar-Saavedra, F.~Deppisch, O.~Kittel and J.~W.~F. Valle, \emph{{Flavour
  in heavy neutrino searches at the LHC}}, {\emph{Phys.Rev.} {\bf D85} (2012)
  091301}, [\href{https://arxiv.org/abs/1203.5998}{{\tt 1203.5998}}].

\bibitem{Das:2012ii}
S.~Das, F.~Deppisch, O.~Kittel and J.~W.~F. Valle, \emph{{Heavy Neutrinos and
  Lepton Flavour Violation in Left-Right Symmetric Models at the LHC}},
  {\emph{Phys.Rev.} {\bf D86} (2012) 055006},
  [\href{https://arxiv.org/abs/1206.0256}{{\tt 1206.0256}}].

\bibitem{Davidson:1987mh}
A.~Davidson and K.~C. Wali, \emph{Universal seesaw mechanism?}, {\emph{Phys.
  Rev. Lett.} {\bf 59} (1987) 393}.

\bibitem{Valencia:1994cj}
G.~Valencia and S.~Willenbrock, \emph{{Quark - lepton unification and rare
  meson decays}}, \href{http://dx.doi.org/10.1103/PhysRevD.50.6843}{\emph{Phys.
  Rev.} {\bf D50} (1994) 6843--6848},
  [\href{https://arxiv.org/abs/hep-ph/9409201}{{\tt hep-ph/9409201}}].

\bibitem{Smirnov:2007hv}
A.~D. Smirnov, \emph{{Mass limits for scalar and gauge leptoquarks from K(L)0
  ---> e-+ mu+-, B0 ---> e-+ tau+- decays}},
  \href{http://dx.doi.org/10.1142/S0217732307024401}{\emph{Mod. Phys. Lett.}
  {\bf A22} (2007) 2353--2363}, [\href{https://arxiv.org/abs/0705.0308}{{\tt
  0705.0308}}].

\bibitem{Hartmann:2014fya}
F.~Hartmann, W.~Kilian and K.~Schnitter, \emph{{Multiple Scales in Pati-Salam
  Unification Models}},
  \href{http://dx.doi.org/10.1007/JHEP05(2014)064}{\emph{JHEP} {\bf 05} (2014)
  064}, [\href{https://arxiv.org/abs/1401.7891}{{\tt 1401.7891}}].

\bibitem{Grimus:2000vj}
W.~Grimus and L.~Lavoura, \emph{The seesaw mechanism at arbitrary order:
  Disentangling the small scale from the large scale}, {\emph{JHEP} {\bf 11}
  (2000) 042}, [\href{https://arxiv.org/abs/hep-ph/0008179}{{\tt
  hep-ph/0008179}}].

\bibitem{Bhattacharyya:2012pi}
G.~Bhattacharyya, I.~de~Medeiros~Varzielas and P.~Leser, \emph{{A common origin
  of fermion mixing and geometrical CP violation, and its test through Higgs
  physics at the LHC}},
  \href{http://dx.doi.org/10.1103/PhysRevLett.109.241603}{\emph{Phys. Rev.
  Lett.} {\bf 109} (2012) 241603}, [\href{https://arxiv.org/abs/1210.0545}{{\tt
  1210.0545}}].

\bibitem{Bjorkeroth:2015uou}
F.~Björkeroth, F.~J. de~Anda, I.~de~Medeiros~Varzielas and S.~F. King,
  \emph{{Towards a complete $\Delta(27) \times SO(10)$ SUSY GUT}},
  \href{http://dx.doi.org/10.1103/PhysRevD.94.016006}{\emph{Phys. Rev.} {\bf
  D94} (2016) 016006}, [\href{https://arxiv.org/abs/1512.00850}{{\tt
  1512.00850}}].

\bibitem{Chen:2015jta}
P.~Chen et~al., \emph{{Warped flavor symmetry predictions for neutrino
  physics}}, \href{http://dx.doi.org/10.1007/JHEP01(2016)007}{\emph{JHEP} {\bf
  01} (2016) 007}, [\href{https://arxiv.org/abs/1509.06683}{{\tt 1509.06683}}].

\bibitem{Vien:2016tmh}
V.~V. Vien, A.~E. C\'arcamo~Hern\'andez and H.~N. Long, \emph{{The $\Delta(27)$
  flavor 3-3-1 model with neutral leptons}},
  \href{http://dx.doi.org/10.1016/j.nuclphysb.2016.10.010}{\emph{Nucl. Phys.}
  {\bf B913} (2016) 792--814}, [\href{https://arxiv.org/abs/1601.03300}{{\tt
  1601.03300}}].

\bibitem{Hernandez:2016eod}
A.~E. Cárcamo~Hernández, H.~N. Long and V.~V. Vien, \emph{{A 3-3-1 model with
  right-handed neutrinos based on the $\varDelta \left( 27\right) $ family
  symmetry}},
  \href{http://dx.doi.org/10.1140/epjc/s10052-016-4074-0}{\emph{Eur. Phys. J.}
  {\bf C76} (2016) 242}, [\href{https://arxiv.org/abs/1601.05062}{{\tt
  1601.05062}}].

\bibitem{Hernandez:2015tna}
A.~E. C\'{a}rcamo~Hern\'{a}ndez and R.~Martinez, \emph{{A predictive 3-3-1
  model with $A_4$ flavor symmetry}},
  \href{http://dx.doi.org/10.1016/j.nuclphysb.2016.02.025}{\emph{Nucl. Phys.}
  {\bf B905} (2016) 337--358}, [\href{https://arxiv.org/abs/1501.05937}{{\tt
  1501.05937}}].

\bibitem{CarcamoHernandez:2017kra}
A.~E. Cárcamo~Hernández and H.~N. Long, \emph{{A highly predictive $A_{4}$
  flavour 3-3-1 model with radiative inverse seesaw mechanism}},
  \href{https://arxiv.org/abs/1705.05246}{{\tt 1705.05246}}.

\bibitem{Hernandez:2013hea}
A.~E. C\'{a}rcamo~Hern\'{a}ndez, R.~Martinez and F.~Ochoa, \emph{{Fermion
  masses and mixings in the 3-3-1 model with right-handed neutrinos based on
  the $S_3$ flavor symmetry}},
  \href{http://dx.doi.org/10.1140/epjc/s10052-016-4480-3}{\emph{Eur. Phys. J.}
  {\bf C76} (2016) 634}, [\href{https://arxiv.org/abs/1309.6567}{{\tt
  1309.6567}}].

\bibitem{Emmanuel-Costa:2013gia}
D.~Emmanuel-Costa, C.~Simoes and M.~Tortola, \emph{{The minimal adjoint-SU(5) x
  $Z_{4}$ GUT model}},
  \href{http://dx.doi.org/10.1007/JHEP10(2013)054}{\emph{JHEP} {\bf 10} (2013)
  054}, [\href{https://arxiv.org/abs/1303.5699}{{\tt 1303.5699}}].

\bibitem{Arbelaez:2015toa}
C.~Arbel\'{a}ez, A.~E. C\'{a}rcamo~Hern\'{a}ndez, S.~Kovalenko and I.~Schmidt,
  \emph{{Adjoint $SU(5)$ GUT model with $T_{7}$ flavor symmetry}},
  \href{http://dx.doi.org/10.1103/PhysRevD.92.115015}{\emph{Phys. Rev.} {\bf
  D92} (2015) 115015}, [\href{https://arxiv.org/abs/1507.03852}{{\tt
  1507.03852}}].

\bibitem{Hernandez:2014vta}
A.~E. C\'{a}rcamo~Hern\'{a}ndez, R.~Martinez and J.~Nisperuza, \emph{{$S_3$
  discrete group as a source of the quark mass and mixing pattern in $331$
  models}}, \href{http://dx.doi.org/10.1140/epjc/s10052-015-3278-z}{\emph{Eur.
  Phys. J.} {\bf C75} (2015) 72}, [\href{https://arxiv.org/abs/1401.0937}{{\tt
  1401.0937}}].

\bibitem{Hernandez:2015zeh}
A.~E. C\'{a}rcamo~Hern\'{a}ndez, I.~de~Medeiros~Varzielas and N.~A. Neill,
  \emph{{Novel Randall-Sundrum model with $S_{3}$ flavor symmetry}},
  \href{http://dx.doi.org/10.1103/PhysRevD.94.033011}{\emph{Phys. Rev.} {\bf
  D94} (2016) 033011}, [\href{https://arxiv.org/abs/1511.07420}{{\tt
  1511.07420}}].

\bibitem{joshipura:1992hp}
A.~S. Joshipura and J.~W.~F. Valle, \emph{{Invisible Higgs decays and neutrino
  physics}}, {\emph{Nucl. Phys.} {\bf B397} (1993) 105--122}.

\bibitem{Bonilla:2015jdf}
C.~Bonilla, J.~C. Romao and J.~W.~F. Valle, \emph{{Electroweak breaking and
  neutrino mass: invisible Higgs decays at the LHC (type II seesaw)}},
  \href{http://dx.doi.org/10.1088/1367-2630/18/3/033033}{\emph{New J. Phys.}
  {\bf 18} (2016) 033033}, [\href{https://arxiv.org/abs/1511.07351}{{\tt
  1511.07351}}].

\bibitem{Bora:2012tx}
K.~Bora, \emph{{Updated values of running quark and lepton masses at GUT scale
  in SM, 2HDM and MSSM}},  \href{https://arxiv.org/abs/1206.5909}{{\tt
  1206.5909}}.

\bibitem{Xing:2007fb}
Z.-z. Xing, H.~Zhang and S.~Zhou, \emph{{Updated Values of Running Quark and
  Lepton Masses}}, {\emph{Phys. Rev.} {\bf D77} (2008) 113016},
  [\href{https://arxiv.org/abs/0712.1419}{{\tt 0712.1419}}].

\bibitem{Olive:2016xmw}
{\scshape Particle Data Group} collaboration, C.~Patrignani et~al.,
  \emph{{Review of Particle Physics}},
  \href{http://dx.doi.org/10.1088/1674-1137/40/10/100001}{\emph{Chin. Phys.}
  {\bf C40} (2016) 100001}.

\bibitem{Bifani:2017gyn}
S.~Bifani, \emph{{Status of New Physics searches with $b \to s
  \ell^{+}\ell^{-}$ transitions @ LHCb}},  2017.
\newblock \href{https://arxiv.org/abs/1705.02693}{{\tt 1705.02693}}.

\bibitem{babu:2002dz}
K.~S. Babu, E.~Ma and J.~W.~F. Valle, \emph{Underlying a(4) symmetry for the
  neutrino mass matrix and the quark mixing matrix}, {\emph{Phys. Lett.} {\bf
  B552} (2003) 207--213}, [\href{https://arxiv.org/abs/hep-ph/0206292}{{\tt
  hep-ph/0206292}}].

\bibitem{grimus:2003yn}
W.~Grimus and L.~Lavoura, \emph{{A non-standard CP transformation leading to
  maximal atmospheric neutrino mixing}}, {\emph{Phys. Lett.} {\bf B579} (2004)
  113--122}, [\href{https://arxiv.org/abs/hep-ph/0305309}{{\tt
  hep-ph/0305309}}].

\bibitem{King:2014nza}
S.~F. King, A.~Merle, S.~Morisi, Y.~Shimizu and M.~Tanimoto, \emph{{Neutrino
  Mass and Mixing: from Theory to Experiment}},
  \href{http://dx.doi.org/10.1088/1367-2630/16/4/045018}{\emph{New J.Phys.}
  {\bf 16} (2014) 045018}, [\href{https://arxiv.org/abs/1402.4271}{{\tt
  1402.4271}}].

\bibitem{Chen:2015siy}
P.~Chen, G.-J. Ding, F.~Gonzalez-Canales and J.~W.~F. Valle, \emph{{Generalized
  $\mu-\tau$ reflection symmetry and leptonic CP violation}},
  \href{http://dx.doi.org/10.1016/j.physletb.2015.12.069}{\emph{Phys. Lett.}
  {\bf B753} (2016) 644--652}, [\href{https://arxiv.org/abs/1512.01551}{{\tt
  1512.01551}}].

\bibitem{Chen:2016ica}
P.~Chen, G.-J. Ding, F.~Gonzalez-Canales and J.~W.~F. Valle, \emph{{Classifying
  CP transformations according to their texture zeros: theory and
  implications}},
  \href{http://dx.doi.org/10.1103/PhysRevD.94.033002}{\emph{Phys. Rev.} {\bf
  D94} (2016) 033002}, [\href{https://arxiv.org/abs/1604.03510}{{\tt
  1604.03510}}].

\bibitem{Schechter:1980gr}
J.~Schechter and J.~W.~F. Valle, \emph{{Neutrino Masses in SU(2) x U(1)
  Theories}},
  \href{http://dx.doi.org/10.1103/PhysRevD.22.2227}{\emph{Phys.Rev.} {\bf D22}
  (1980) 2227}.

\bibitem{Rodejohann:2011vc}
W.~Rodejohann and J.~W.~F. Valle, \emph{{Symmetrical Parametrizations of the
  Lepton Mixing Matrix}}, {\emph{Phys.Rev.} {\bf D84} (2011) 073011},
  [\href{https://arxiv.org/abs/1108.3484}{{\tt 1108.3484}}].

\bibitem{Forero:2014bxa}
D.~Forero, M.~Tortola and J.~W.~F. Valle, \emph{{Neutrino oscillations
  refitted}},
  \href{http://dx.doi.org/10.1103/PhysRevD.90.093006}{\emph{Phys.Rev.} {\bf
  D90} (2014) 093006}, [\href{https://arxiv.org/abs/1405.7540}{{\tt
  1405.7540}}].

\bibitem{Alessandria:2011rc}
F.~Alessandria et~al., \emph{{Sensitivity of CUORE to Neutrinoless Double-Beta
  Decay}},  \href{https://arxiv.org/abs/1109.0494}{{\tt 1109.0494}}.

\bibitem{KamLAND-Zen:2016pfg}
{\scshape KamLAND-Zen} collaboration, A.~Gando et~al., \emph{{Search for
  Majorana Neutrinos near the Inverted Mass Hierarchy Region with
  KamLAND-Zen}}, \href{http://dx.doi.org/10.1103/PhysRevLett.117.109903,
  10.1103/PhysRevLett.117.082503}{\emph{Phys. Rev. Lett.} {\bf 117} (2016)
  082503}, [\href{https://arxiv.org/abs/1605.02889}{{\tt 1605.02889}}].

\bibitem{Abt:2004yk}
I.~Abt et~al., \emph{A new ge-76 double beta decay experiment at lngs},
  \href{https://arxiv.org/abs/hep-ex/0404039}{{\tt hep-ex/0404039}}.

\bibitem{Ackermann:2012xja}
{\scshape GERDA} collaboration, K.~H. Ackermann et~al., \emph{{The GERDA
  experiment for the search of $0\nu\beta\beta$ decay in $^{76}$Ge}},
  \href{http://dx.doi.org/10.1140/epjc/s10052-013-2330-0}{\emph{Eur. Phys. J.}
  {\bf C73} (2013) 2330}, [\href{https://arxiv.org/abs/1212.4067}{{\tt
  1212.4067}}].

\bibitem{KamLANDZen:2012aa}
{\scshape KamLAND-Zen Collaboration} collaboration, A.~Gando et~al.,
  \emph{{Measurement of the double beta decay half-life of $^{136}$Xe with the
  KamLAND-Zen experiment}}, {\emph{Phys.Rev.} {\bf C85} (2012) 045504},
  [\href{https://arxiv.org/abs/1201.4664}{{\tt 1201.4664}}].

\bibitem{Albert:2014fya}
{\scshape EXO-200} collaboration, J.~B. Albert et~al., \emph{{Search for
  Majoron-emitting modes of double-beta decay of $^{136}$Xe with EXO-200}},
  \href{http://dx.doi.org/10.1103/PhysRevD.90.092004}{\emph{Phys. Rev.} {\bf
  D90} (2014) 092004}, [\href{https://arxiv.org/abs/1409.6829}{{\tt
  1409.6829}}].

\bibitem{Guiseppe:2011me}
{\scshape Majorana} collaboration, C.~E. Aalseth et~al., \emph{{The Majorana
  Experiment}},
  \href{http://dx.doi.org/10.1016/j.nuclphysbps.2011.04.063}{\emph{Nucl. Phys.
  Proc. Suppl.} {\bf 217} (2011) 44--46},
  [\href{https://arxiv.org/abs/1101.0119}{{\tt 1101.0119}}].

\bibitem{Dorame:2011eb}
L.~Dorame et~al., \emph{{Constraining Neutrinoless Double Beta Decay}},
  {\emph{Nucl.Phys.} {\bf B861} (2012) 259--270},
  [\href{https://arxiv.org/abs/1111.5614}{{\tt 1111.5614}}].

\bibitem{Dorame:2012zv}
L.~Dorame et~al., \emph{{A new neutrino mass sum rule from inverse seesaw}},
  {\emph{Phys.Rev.} {\bf D86} (2012) 056001},
  [\href{https://arxiv.org/abs/1203.0155}{{\tt 1203.0155}}].

\bibitem{King:2013hj}
S.~King, S.~Morisi, E.~Peinado and J.~W.~F. Valle, \emph{{Quark-Lepton Mass
  Relation in a Realistic A4 Extension of the Standard Model}}, {\emph{Phys.
  Lett. B} {\bf 724} (2013) 68--72},
  [\href{https://arxiv.org/abs/1301.7065}{{\tt 1301.7065}}].

\bibitem{Bonilla:2014xla}
C.~Bonilla, S.~Morisi, E.~Peinado and J.~W.~F. Valle, \emph{{Relating quarks
  and leptons with the $T_7$ flavour group}},
  \href{http://dx.doi.org/10.1016/j.physletb.2015.01.017}{\emph{Phys. Lett.}
  {\bf B742} (2015) 99--106}, [\href{https://arxiv.org/abs/1411.4883}{{\tt
  1411.4883}}].

\bibitem{Gehrlein:2016wlc}
J.~Gehrlein, A.~Merle and M.~Spinrath, \emph{{Predictivity of Neutrino Mass Sum
  Rules}}, \href{http://dx.doi.org/10.1103/PhysRevD.94.093003}{\emph{Phys.
  Rev.} {\bf D94} (2016) 093003}, [\href{https://arxiv.org/abs/1606.04965}{{\tt
  1606.04965}}].

\bibitem{Queiroz:2016gif}
F.~S. Queiroz, C.~Siqueira and J.~W.~F. Valle, \emph{{Constraining Flavor
  Changing Interactions from LHC Run-2 Dilepton Bounds with Vector Mediators}},
  \href{http://dx.doi.org/10.1016/j.physletb.2016.10.057}{\emph{Phys. Lett.}
  {\bf B763} (2016) 269--274}, [\href{https://arxiv.org/abs/1608.07295}{{\tt
  1608.07295}}].

\end{thebibliography}\endgroup
%BIBLIOGRAPHY FILE HAS TO BE IN BIB EXTENSION.

\end{document}